\pdfoutput=1
\documentclass[acmsmall,screen]{acmart}

\AtBeginDocument{%
  }

\setcopyright{acmlicensed}
\copyrightyear{2025}
\acmYear{2025}
\acmDOI{XXXXXXX.XXXXXXX}

\acmJournal{JACM}
\acmVolume{37}
\acmNumber{4}
\acmArticle{111}
\acmMonth{8}

\usepackage{threeparttable} 
\usepackage{booktabs}
\usepackage{multirow} 
\usepackage{subfigure}
\begin{document}

\title{MMHCL: Multi-Modal Hypergraph  Contrastive Learning for Recommendation}

\author{Xu Guo}
\email{xu.guo@njust.edu.cn}
\author{Tong Zhang}
\email{tong.zhang@njust.edu.cn}
\author{Fuyun Wang}
\email{fyw271828@njust.edu.cn}
\author{Xudong Wang}
\email{xd_wang@njust.edu.cn}
\author{Xiaoya Zhang}
\email{zhangxiaoya@njust.edu.cn}
\affiliation{%
  \institution{School of Computer Science and Engineering, Nanjing University of Science and Technology}
  \city{Nanjing}
  \country{China}
}

\author{Xin Liu}
\affiliation{%
  \institution{Shituoyun (Nanjing) Technology Co., Ltd}
  \city{Nanjing}
  \country{China}}
\email{xin.liu@seetacloud.com}

\author{Zhen Cui}
\affiliation{%
  \institution{School of Artificial Intelligence, Beijing Normal University}
  \city{Beijing}
  \country{China}}
\email{zhen.cui@bnu.edu.cn}
\authornote{Corresponding author: Zhen Cui.}

\renewcommand{\shortauthors}{Xu Guo, Tong Zhang, Zhen Cui, et al.}

\begin{abstract}
    The burgeoning presence of multimodal content-sharing platforms propels the development of personalized recommender systems. 
    Previous works usually suffer from data sparsity and cold-start problems, and may fail to adequately explore semantic user-product associations from multimodal data. 
    To address these issues, we propose a novel Multi-Modal Hypergraph Contrastive Learning (MMHCL) framework for user recommendation. 
    For a comprehensive information exploration from user-product relations, we construct two hypergraphs, i.e. a user-to-user (u2u) hypergraph and an item-to-item (i2i) hypergraph, to mine shared preferences among users and intricate multimodal semantic resemblance among items, respectively. 
    This process yields denser second-order semantics that are fused with first-order user-item interaction as complementary to alleviate the data sparsity issue. 
    Then, we design a contrastive feature enhancement paradigm by applying synergistic contrastive learning. By maximizing/minimizing the mutual information between second-order (e.g. shared preference pattern for users) and first-order (information of selected items for users) embeddings of the same/different users and items, the feature distinguishability can be effectively enhanced. 
    Compared with using sparse primary user-item interaction only, our MMHCL obtains denser second-order hypergraphs and excavates more abundant shared attributes to explore the user-product associations, which to a certain extent alleviates the problems of data sparsity and cold-start. 
    Extensive experiments have comprehensively demonstrated the effectiveness of our method. 
    Our code is publicly available at \url{https://github.com/Xu107/MMHCL}. 
\end{abstract}




\maketitle

\section{Introduction}
    Recommender systems play an essential role in alleviating information overload while mining users’ implicit requirements and preferences ~\cite{he2020lightgcn,ying2018graph,liu2024graph,liu2024inter,wen2024cdcm,liu2024privacy}. 
    In recent years, the research interest in multimodal recommendation, which integrates diverse modalities of item content (e.g., visual, acoustic, and textual), has been steadily increasing ~\cite{wei2019mmgcn,zhang2021mining,tao2022self,wei2023multi,zhou2023bootstrap,zhou2023tale,guo2025m}. 

    Previous research predominantly falls into two main categories: graph neural networks (GNNs)-based methods \cite{wu2022graph,wang2019neural} and more recent approaches incorporating self-supervised learning (SSL)~\cite{yu2023self,wu2021self}. 
    GNNs-based methods~\cite{he2020lightgcn,fan2019graph,shen2021powerful}  model large-scale recommendation data as graph structures, and leverage advanced graph neural networks ~\cite{kipf2016semi, hamilton2017inductive, velivckovic2017graph} to uncover intricate higher-order relationships among users and items while effectively capturing collaborative filtering signals. 
    For instance, models such as MMGCN \cite{wei2019mmgcn} and GRCN \cite{wei2020graph} generate user-item interaction bipartite graphs for different modalities, leveraging graph convolution to predict user preferences within each modality. 
    Despite their success, in practice, users generally tend to interact with only a small subset of available items \cite{yu2023self}. 
    This results in sparse user-item interactions, which are insufficient for strong supervisory signals, exacerbating issues such as data sparsity and cold-start problems. 

    Self-supervised learning (SSL) ~\cite{liu2021self,yu2023self,ren2024sslrec,cao2025enhancing} gives natural echoes to the data sparsity problem by deriving supervision signals directly from the data itself, effectively reducing the dependence on scarce labeled data. 
    Among SSL techniques, contrastive learning (CL) \cite{khosla2020supervised,wang2022contrastive} is particularly prominent. 
    It aims to enhance the discriminability of data representations by minimizing the distance between similar instances and maximizing the distance between dissimilar ones in the representation space. 
    Several CL-based methods have demonstrated superiority in ID-based collaborative filtering frameworks ~\cite{wu2021self,yu2022graph,jing2023contrastive,yu2023xsimgcl}. 
    However, the application of CL in multimodal recommendation remains insufficiently explored. 
    Existing methods such as SLMRec \cite{tao2022self} and MMGCL \cite{yi2022multi} generate multiple views for CL by applying dropout and masking operations at the feature and modality levels, respectively. 
    While effective, these operations may inadvertently discard valuable information from raw data, and excessive dropout ratios lead to multimodal feature collapse. 
    Additionally, they may overlook the rich higher-order semantic structures inherent in both user and item levels. 

    Despite their effectiveness, previous methods encounter the following challenges:
    i) Data sparsity: Observed user-item interactions are usually extremely sparse across the entire interaction space. Methods that rely on modeling first-order interactions may fail to extract adequate supervisory signals. 
    ii) Insufficient feature distinguishability: Existing collaborative filtering methods focus on mining interaction patterns between users and items based on historical behavior records, while paying less attention to the distinguishability of features among users/items. 
    iii) Inadequate multimodal exploration: The structural correlations among items across different modalities are crucial for capturing item characteristics but are often insufficiently modeled in existing approaches. 
 
\begin{figure}[t]
    \centering
    \includegraphics[width=0.75\linewidth]{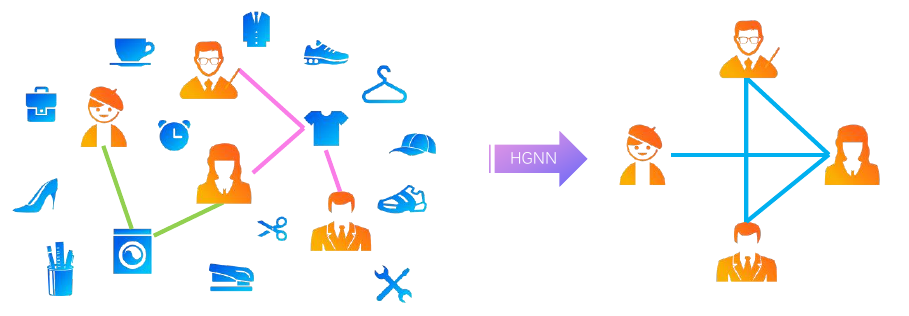}
    \caption{
    Example of recommendation. 
    We introduce a Hypergraph Neural Network (HGNN) for explicitly model shared preferences among users. 
    The left side illustrates a complex user-item interaction graph where different colors indicate distinct hyperedges, each connecting multiple users and items. 
    On the right, HGNN aggregates these second-order relationships to explicitly uncover deeper semantic connections, extracting denser, higher-order information.}
    
    \label{intro}
    \end{figure}

    To address these challenges, we introduce multimodal hypergraph learning to enhance recommendation performance. 
    Our approach is motivated by the observation that users following popular trends are more likely to choose trending items. 
    This user behavior allows us to derive an item's "hotness index" from a user perspective. 
    Therefore, we design hypergraphs to explicitly model intricate second-order information of relationships (e.g., shared user preferences among users) and utilize them to complement first-order user-item interactions, alleviating data sparsity problems. 
    The hypergraph \cite{feng2019hypergraph,yadati2019hypergcn,gao2022hgnn+} consists of nodes (i.e., users/items) and hyperedges, with each hyperedge can connect several nodes. 
    Figure.\ref{intro} illustrates a toy example of user hypergraph construction, where nodes represent users/items, and hyperedges connect multiple nodes. 
    As shown, denser second-order user-user correlations are obtained by using the user-item connection as a bridge. 
    Moreover, the multimodal characteristic of items, with intricate associations across modalities, encourages us to extract comprehensive structural correlations to enhance feature representation. 
    
    Building on the above analysis, we propose a novel framework, Multimodal Hypergraph Contrastive Learning (MMHCL), designed to comprehensively capture shared user preferences and intricate higher-order multimodal semantic information for items. 
    To thoroughly explore user-product relationships in depth, in addition to primary first-order user-item interactions, we construct two additional user-to-user (u2u) and item-to-item (i2i) hypergraphs. 
    These hypergraphs effectively reveal second-order semantic relationships among users, such as shared preferences, while capturing intricate multimodal similarities between items. This enables a more comprehensive and nuanced excavation of user-item interaction information, enriching the overall recommendation process. 
    Specifically, the u2u hypergraph explicitly models user-user relationships to integrate second-order preference information, enriching the understanding of shared user behaviors. 
    For multimodal fusion and complementarity, we design the i2i hypergraph to not only facilitate intra-modal interactions but also enable inter-modal fusion, uncovering potentially intricate higher-order semantic connections between items from various modalities. 
    This process produces denser second-order semantics, which are fused with first-order user-item interactions to complement and alleviate the data sparsity issue. 
    To further enhance feature distinguishability, we introduce a contrastive learning framework, applying synergistic contrastive learning between first-order and second-order embeddings of users/items. 
    Contrastive learning maximizes/minimizes the mutual information between second-order and first-order embeddings of the same or different users and items. 
    Compared to approaches that rely solely on sparse first-order interactions, our MMHCL constructs denser second-order hypergraphs and excavates more abundant shared attributes to explore the user-item association, which to a certain extent alleviates the problems of data sparsity and cold-start. 
    We evaluate the proposed approach on three widely used multimodal recommendation datasets, and the experimental results demonstrate the effectiveness of our methods.

    The contributions of this work can be outlined as follows: 

    \begin{itemize}
    \item{
    We propose a novel and effective MMHCL framework for the multimodal recommendation that leverages hypergraphs to explore intricate higher-order shared preferences among users and potential multimodal semantic correlations for items. 
    }
    \item{
    We introduce a contrastive feature enhancement framework tailored for the multimodal recommendation, using hypergraphs to learn representations at both user and item levels, enabling comprehensive synergistic contrastive learning. 
    }
    \item{
    Extensive experiments on three public datasets robustly validate the superiority of our proposed model over various baselines. 
    }
\end{itemize}

\textbf{Organizational Structure of the Paper}

To facilitate a clear understanding of our approach, the rest of the paper is structured as follows:

\begin{itemize}
    \item Section 2 (Related Work): This section presents a comprehensive review of related research, covering graph neural networks, self-supervised learning techniques, and hypergraph-based approaches in recommendation systems.
    \item Section 3 (The Proposed Method): This section begins with the formal notation  of notation and the recommendation task, followed by an overview of the proposed framework. We then describe the construction of user-to-user (u2u) and item-to-item (i2i) hypergraphs, designed to capture higher-order shared user preferences and multimodal item correlations, respectively. Next, we detail the model fusion strategies and introduce our synergistic contrastive learning mechanism. Finally, we present the loss function for model optimization. 
    \item Section 4 (Experiments): This section evaluates the effectiveness of MMHCL through extensive experiments on three multimodal recommendation datasets, comparing against state-of-the-art baselines. Analyzes the contributions of different components of MMHCL. Furthermore, we conduct a hyperparameter sensitivity study, exploring the effects of key factors such as contrastive learning temperature, graph convolution depth, and embedding dimensionality. Finally, the influence of different modalities is explored. 
    \item Section 5 (Conclusion): This section summarizes and discusses our proposed framework.

\end{itemize}

\section{Related Work}
\subsection{GNN-based Recommender Systems}

    Graph neural networks (GNN) are extensively used in recommender systems to model graph-structured data and learn representations ~\cite{wu2022graph}. 
    These approaches can be grouped into five categories: 
    1) Collaborative filtering based on user-item interactions (e.g., NGCF~\cite{wang2019neural}, LightGCN~\cite{he2020lightgcn}). 
    2) Sequential and session recommendation based on time-series user historical behaviors (e.g., SR-GNN \cite{wu2019session}, GC-SAN \cite{xu2019graph}, FGNN \cite{qiu2019rethinking}, DHCN \cite{xia2021self}, CGSNet \cite{wang2022cgsnet}, CM-GNN \cite{wang2023contrastive}, MGs \cite{lai2022attribute}, S$^{3}$-Rec \cite{zhou2020s3}). 
    3) Social recommendation based on user associations (e.g., GraphRec \cite{fan2019graph}, ESRF \cite{yu2020enhancing}).  
    4) Knowledge graph based recommendation (e.g., KGCN \cite{wang2019knowledge}, KGAT \cite{wang2019kgat}, KGRec \cite{yang2023knowledge}). 
    5) Multimodal based recommendation ~\cite{he2016vbpr,wei2019mmgcn,wei2020graph,zhang2021mining,wang2021dualgnn,yi2021multi,zhou2023tale,li2024hypergraph}. 
    In the multimodal context, early work such as VBPR \cite{he2016vbpr} integrates item ID embeddings with visual features for matrix factorization. 
    ACF \cite{chen2017attentive} and VECF \cite{chen2019personalized} leverage hierarchical attention mechanisms to capture complex specific modality preferences of users. 
    MMGCN \cite{wei2019mmgcn} pioneered and introduced graph convolution to the multimodal recommendation by applying it to each modality. 
    GRCN \cite{wei2020graph} identifies false-positive feedback and removes noisy edges from the interaction graph. 
    LATTICE \cite{zhang2021mining} uses deep neural networks to learn latent items' structure information. 
    FREEDOM \cite{zhou2023tale} freezes LATTICE learnable items' structure and refines this by denoising user-item interactions. 
    Some methods exhibit a strong reliance on user-item interactions and confront challenges related to data sparsity. 

\subsection{Self-Supervised Learning for Recommendation}
    Self-supervised learning (SSL)~\cite{bachman2019learning,he2020momentum,liu2021self} utilizes auxiliary learning tasks to augment the original supervised signals to enhance the performance of recommendation systems, without relying solely on explicit user-item interactions data \cite{yu2023self}. 
    In collaborative filtering, SGL \cite{wu2021self} and SimGCL \cite{yu2022graph} conduct data augmentation by perturbing the graph structure and adding random noise to the embeddings, respectively. 
    In sequential recommendation, CL4SRec \cite{xie2022contrastive} expands the user's historical sequence by cropping, masking, and reordering. 
    ICLRec \cite{chen2022intent} performs contrastive learning between a behavior sequence and its corresponding intent prototype. 
    In the knowledge graph-based recommendation, MCCLK \cite{zou2022multi} considers different views for contrastive learning and so on. 
    In multimodal recommendation, SLMRec \cite{tao2022self} and MMGCL \cite{yi2022multi} generate various representation views by using dropout and mask operations at the feature and modality levels, respectively. 
    MMSSL \cite{wei2023multi} generates modality-aware interactive structures through adversarial perturbations and applies contrastive learning across modalities. 
    BM3 \cite{zhou2023bootstrap} applies a siamese neural network to generate diverse views of the features for data augmentation. 

\subsection{Hypergraph Learning for Recommendation}
    Hypergraph neural network (HGNN) ~\cite{feng2019hypergraph,yadati2019hypergcn} naturally captures intricate higher-order semantics among nodes and has recently garnered attention in recommender systems. 
    In contrast to a simple graph, a hyperedge in a hypergraph connects two or more vertices ~\cite{feng2019hypergraph,yadati2019hypergcn}. 
    In collaborative filtering recommendation, HCCF \cite{xia2022hypergraph} constructs a learnable hypergraph to encapsulate both local and global collaborative relationships. 
    In sequential and session recommendation, researchers have made various attempts to explore the capability of hypergraphs to model temporal data ~\cite{xia2021self,li2021hyperbolic,yang2022multi,wang2021session,wang2020next,li2022enhancing}. 
    HyRec \cite{wang2020next} employs hypergraphs to represent short-term user preferences for the sequential recommendation. 
    DHCF \cite{xia2021self} constructs a dual-channel hypergraph comprising session-based hypergraph and line graph to maximize the mutual information between the session representations learned. 
    HEML \cite{li2024hypergraph} employs dual-scale transformers, capsule networks, and hypergraph convolutional networks to model multi-dimensional user interests for enhanced multi-behavior sequential recommendations. 
    In social recommendation, MHCN \cite{yu2021self} encodes a hypergraph for each channel in the network. 
    In knowledge graph-based recommendation, SDK \cite{liu2023self} introduces a cross-view dynamic hypergraph self-supervised learning framework to enhance knowledge graphs. LGMRec\cite{guo2024lgmrec} integrates local and global embedding techniques through hypergraph structures. 
    

\section{The Proposed Method}
\subsection{Preliminaries}
\begin{table}[h]
\centering
\caption{Notation Summary}
\begin{tabular}{|l|p{10cm}|}
\hline
\textbf{Notation} & \textbf{Description} \\ \hline
$\mathcal{U}$, $\mathcal{I}$ & Set of users and items \\ \hline
$\mathbf{A}$ & User-item interaction matrix ($M \times N$), where $A_{u,i}=1$ if user $u$ interacted with item $i$ (and $0$ otherwise) \\ \hline
$\mathcal{M}$ & Set of item's content modalities (e.g., visual, textual, acoustic) \\ \hline
$\mathbf{e}_{u}$ , $\mathbf{e}_{i}$  & ID embedding vector of user $u$ / item $i$ (learned from the collaborative filtering backbone) \\ \hline
$\mathbf{h}_{u}$ , $\mathbf{h}_{i}$  & Hypergraph-based embedding of user $u$ / item $i$ (learned from u2u / i2i hypergraph convolution layers) \\ \hline
$\widetilde{\mathbf{e}}_u$, $\widetilde{\mathbf{e}}_i$ & Final fused embedding for user $u$ / item $i$ (combining the ID embedding and hypergraph embedding) \\ \hline
$\mathbf{W}_{u2u}$, $\mathbf{W}_{i2i}$ & Diagonal weight matrices for hyperedges in the u2u and i2i hypergraphs (set to identity in our implementation) \\ \hline
$\tau$ & Temperature parameter for contrastive learning loss \\ \hline
$\alpha$, $\beta$ & Loss weight coefficients for the u2u and i2i contrastive terms, respectively \\ \hline
$\lambda$ & Coefficient for $L_2$ regularization (weight decay) \\ \hline

\end{tabular}
\label{Table:0}
\end{table}
    
    Denote $\mathcal{U}$ and $\mathcal{I}$ as the sets containing $M$ users and $N$ items, respectively. The matrix $\mathbf{A} \in \mathbb{R}^{M \times N}$ indicates historical interaction relationships between users and items, with $A_{u,i}=1$ signifying the presence of an interaction record $({u},{i})$ \footnote{For simplification, we mix the use of $u$ and $i$, as one user/item or matrix/vector indices, in clear context.} for $ {u} \in \mathcal{U}$ and ${i} \in \mathcal{I}$, and 0 otherwise. 
    In contexts involving multiple modalities, items are characterized by various multimodal features. We introduce $\mathcal{M}=\{\mathbf{v},\mathbf{t},\mathbf{a}\}$ to represent the set consisting of visual, textual, and acoustic modalities. 
    The objective is to learn a function that estimates the likelihood of a target user $u$ interacting with candidate items, and ultimately to rank or predict the top-$K$ items that $u$ would prefer, given the user-item interaction history and the items' multimodal features. The main notations used in our model are summarized in Table.\ref{Table:0}.

\begin{figure*}
\centering
\includegraphics[width=1\textwidth]{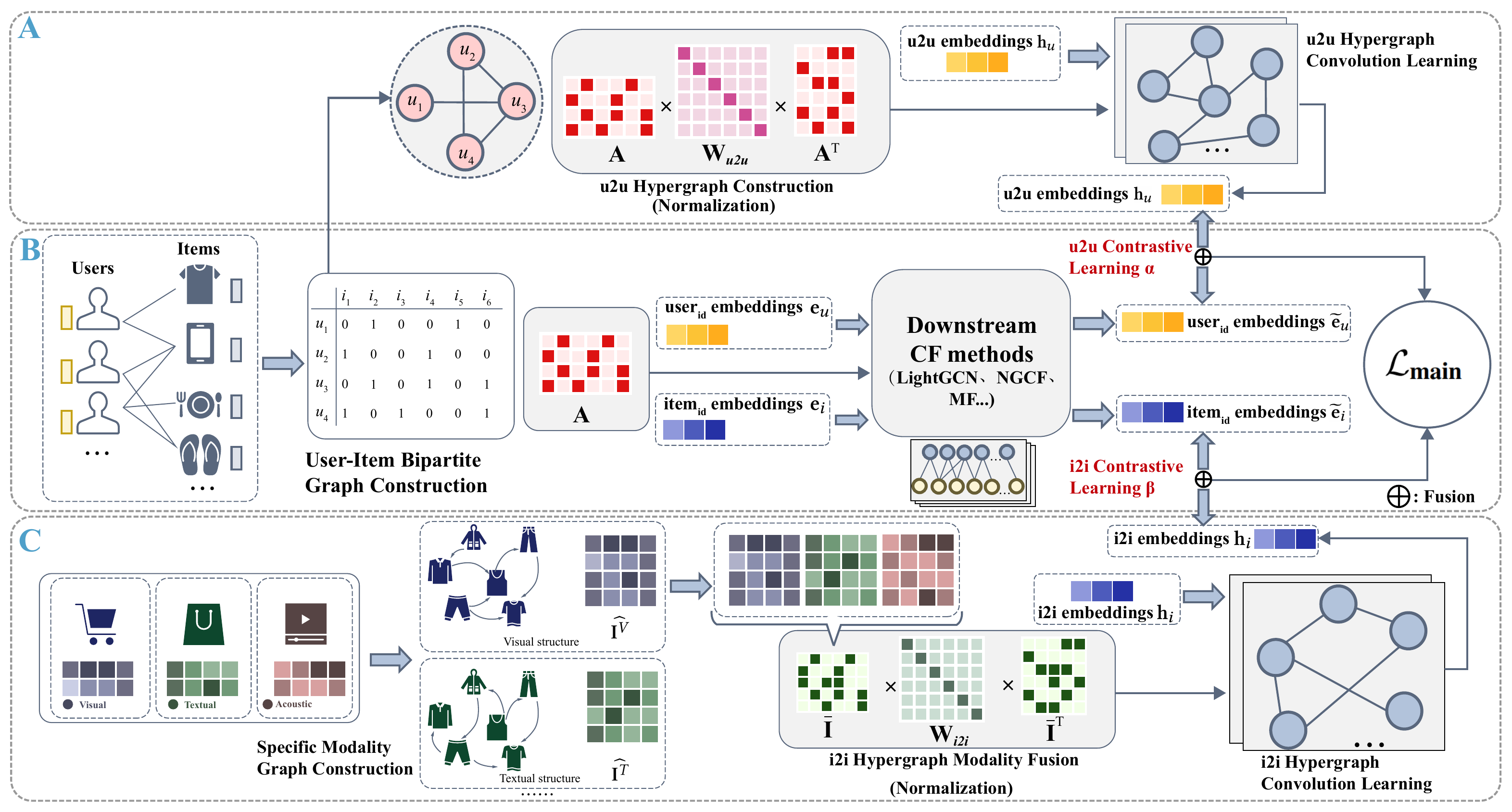}
\hspace{0.2cm}
\vspace{-0.5cm}
\caption{
The structure overview of the proposed MMHCL. 
A) The u2u hypergraph is constructed based on the user-item interactions for learning the user-level representations. 
B) The user/item embeddings acquired from the user/item level are fused with the embeddings learned from the downstream tasks. Meanwhile, synergistic contrastive learning is implemented. 
C) The i2i hypergraph is built using the raw multimodal features of items to learn the item-level representations. 
See the Overview section for details. 
}
\label{fig:model}
\end{figure*}
\subsection{Overview} 
    Fig.\ref{fig:model} shows the whole architecture of our MMHCL framework. 
    The inputs of the MMHCL are user-item interactions and raw multimodal features of items. 
    The model consists of four main components: 
    \textbf{1) user-to-user (u2u) Hypergraph.} In part A of the figure, a u2u hypergraph is generated from user-item interactions to extract higher-order shared preferences among users and learn user-level representations. 
    \textbf{2) item-to-item (i2i) Hypergraph.} In part C of the figure, specific-modal graph structures are initially established based on the raw feature similarity of the items. 
    Subsequently, these unimodal graphs are amalgamated into the i2i hypergraph to concretize and excavate intricate higher-order multimodal semantic associations among items, facilitating the learning of item-level representations. 
    \textbf{3) Fusion with downstream tasks.} On the right side of Part B in the figure, the item- and user-level representations, learned through hypergraph convolution at both ends, are fused with the downstream task representations to provide supplementary information. 
    \textbf{4) Synergistic contrastive learning.} Meanwhile, macro-contrastive learning is performed synergistically at both ends to provide auxiliary supervised signals. 
    Next, we elaborate on the four components. 

\subsection{User-to-user (u2u) Hypergraph} 
    For the user level, the u2u hypergraph is constructed to capture complex shared interest preferences among users. 
    This u2u hypergraph represents users as nodes and their interaction history with items as hyperedges. 
    It enables the exploration of second-order semantic information among users through hypergraph convolution operations, utilizing items as bridges.    
    To construct the hypergraph encoder, we design the hypergraph convolutional operation refer to HGNN \cite{feng2019hypergraph} and stack them to obtain the powerful representation on the user level, which is formulated as: 
    \begin{equation}
        \mathbf{u}_{i}^{(l+1)}=\frac{1}{N} \sum_{j=1}^{M} \sum_{\epsilon=1}^{N} \mathbf{A}_{i \epsilon} \mathbf{A}_{j \epsilon} \mathbf{W}_{\epsilon \epsilon} \mathbf{u}_{j}^{(l)}, 
    \end{equation}
    where $\mathbf{u}_{i}$ denotes the embedding of the $i$-th user $\mathbf{u}$. 
    The $\mathbf{W}_{\epsilon \epsilon}$ is the trainable hyperedges weight and it can be replaced with a predefined user-to-user similarity. 
    Here we set it to 1 for simplicity. 
    We represent our hypergraph convolution in the form of matrix multiplication: 
    \begin{align}
        \overline{\mathbf{H}}_{u2u}&=\mathbf{LN} \left( \mathbf{A} \mathbf{W}_{u2u} \mathbf{A}^{\mathrm{T}} \right), \\
        \mathbf{H}_{u}^{(l+1)}&=\overline{\mathbf{H}}_{u2u} \mathbf{H}_{u}^{(l)},
    \end{align}
    where $\overline{\mathbf{H}}_{u2u}$ is the reorganized form of the concise u2u hypergraph after convolution and $\mathbf{W}_{u2u}$ set as an identity matrix  for simplicity. 
    $\mathbf{H}_{u} \in \mathbb{R}^{M \times d}$ is the matrix form of the user embeddings with Xavier initialization \cite{glorot2010understanding} on the user-level. 
    $\mathbf{LN}$ denotes the layer normalization (e.g., laplacian normalization) is applied to mitigate the risk of gradient explosion during training. 
    To streamline the embedding learning process, we discard the nonlinear activation function. 
    Hypergraphs can be conceptualized as a two-step information transfer process: first from nodes to hyperedges, and then from hyperedges back to nodes. 
    After combining Eq.(2) and Eq.(3), $\mathbf{A}^{\mathrm{T}} \mathbf{H}_{u}^{(l)}$ aggregates user nodes information into the item hyperedges, preceded by matrix multiplication with $\mathbf{A}$, which in turn aggregates item hyperedges information back to the user nodes. 
    From a broader perspective, in the execution of $\mathbf{A}\mathbf{A}^{\mathrm{T}}$, each element within the resultant matrix signifies the count of items between users. 
    Item hyperedges serve as intermediate bridges in the construction of u2u hypergraph, facilitating the extraction of potential intricate high-order shared interest preferences among users after $l$-layers hypergraph convolution. 
    The result of u2u hypergraph convolution after $L$ layers, denoted as $\mathbf{h}_{u}=\mathbf{u}^{(L)}$ , is utilized as the final embeddings at the user level. 

\subsection{Item-to-item (i2i) Hypergraph} 
    For the item level, the i2i hypergraph is constructed to extract potential intricate higher-order semantic correlations among items from multimodal features. 
    Initially, we measure the similarity among items within a specific modality to identify related item neighbors. 
    For clarity, we employ cosine similarity to evaluate the similarity between items within a given modality:
    \begin{equation}
        \mathbf{I}_{ij}^m=\frac{({\mathbf{e}}_i^m)^\top{\mathbf{e}}_j^m}{\|{\mathbf{e}}_i^m\|\|{\mathbf{e}}_j^m\|},
    \end{equation}
    where $\mathbf{I}^m \in \mathbb{R}^{N \times N}$ denotes the correlation matrix that encapsulates the dense and interconnected relationships among items within an unimodality. 
    Due to the dense nature and excessive redundant edge information, We adopt the K-nearest neighbor (KNN) approach to filter out extraneous information, converting it into a sparse matrix for efficiency, which can be expressed as: 
    \begin{equation}
        \widehat{\mathbf{I}_{ij}^m}=\begin{cases}1,\quad \mathbf{I}_{ij}^m\in\text{top-}K(\mathbf{I}_{i}^m),\\0,\quad\text{otherwise}.\end{cases}
    \end{equation}
    
    Then we amalgamate the unimodal information into the hypergraph $\overline{\mathbf{I}}$, which can be represented as: 
    \begin{equation}
            \overline{\mathbf{I}}={||}_{m}^{M} \widehat{\mathbf{I}^m}
    \end{equation}
    where $||$ denotes the concatenation operation. 
    The i2i hypergraph takes the items after fusing the multimodal information as nodes and takes the feature nearest neighbor records of each node in each modality as hyperedges. 
    Similar to the user side, we define i2i hypergraph convolution as follows: 
    \begin{align}
        \overline{\mathbf{H}}_{i2i}&=\mathbf{LN} \left( \overline{\mathbf{I}} \mathbf{W}_{i2i} \overline{\mathbf{I}}^{\mathrm{T}} \right), \\
        \mathbf{H}_{i}^{(l+1)}&=\overline{\mathbf{H}}_{i2i} \mathbf{H}_{i}^{(l)},
    \end{align}
    where $\overline{\mathbf{H}}_{i2i}$ is the reorganization form of i2i hypergraph. 
    The weight $\mathbf{W}_{i2i}$ of the hyperedges between items can be trained or predefined. 
    Here, we simplify it to an identity matrix. 
    The $\mathbf{LN}$ is applied to prevent gradient explosion during training, similar to its use on the user side. 
    $\mathbf{H}_{i}$ is the matrix form of the item embeddings with Xavier initialization on the item level.     
    By doing so, we can transform multimodal data of items into a hypergraph structure, effectively concretizing the intricate higher-order connections among items. 
    After combining Eq.(7) and Eq.(8), $\overline{\mathbf{I}}^{\mathrm{T}}\mathbf{H}_{i}^{(i)}$ aggregates modality-specific item nodes information into hyperedges, promoting intra-modal interaction. 
    The preceding multiplication by $\overline{\mathbf{I}}$ reaggregates the hyperedge information, which connects unimodal features to the item nodes by linking multiple hyperedges, facilitating inter-modal fusion. 
    Stacking multiple i2i hypergraph convolutional layers facilitates intra-modal interaction and the fusion of inter-modal information. 
    This approach extracts intricate higher-order semantic correlations among items and enables effective multimodal fusion. 
    Similarly, the result of i2i hypergraph convolution $\mathbf{h}_{i}$  after $L$-layers is utilized as the final embeddings at the item level.
    
\subsection{Fusion with Downstream Tasks} 
    The two hypergraphs remain largely decoupled from downstream collaborative filtering tasks, and we utilize LightGCN \cite{he2020lightgcn} as the downstream backbone due to its superior generalizability. 
    Following two hypergraph convolutions, we fuse the normalized representations of user level  $\mathbf{h}_{u}$ and item level $\mathbf{h}_{i}$  into the ID-representations $\mathbf{e}_{u}$ and $\mathbf{e}_{i}$  obtained from the downstream collaborative filtering task. 
    This fusion complements the denser second-order information for $\mathbf{e}_{u}$ and $\mathbf{e}_{i}$: 
    \begin{align}
        \widetilde{\mathbf{e}}_u&=\mathbf{e}_u+\frac{\mathbf{h}_u}{\| \mathbf{h}_u\|_2}, \label{eq9} \\ 
        \widetilde{\mathbf{e}}_i&=\mathbf{e}_i+\frac{\mathbf{h}_i}{\| \mathbf{h}_i\|_2},       
        \label{eq10}
    \end{align}
    where $\widetilde{\mathbf{e}}_u$ and $\widetilde{\mathbf{e}}_i$ correspond to the user and item embeddings from the ultimate output of our model. 
    The fusion process integrates additional shared user preference attributes into the user ID embeddings and incorporates multimodal complex higher-order semantic resemblance of items into the item ID embeddings for downstream tasks.
    Meanwhile, the i2i hypergraph circumvents the need for user-item interactions, and effectively alleviates the cold start problem. 
    Following the approaches \cite{wei2023multi}, we utilize the inner product to predict the probability of user $u$ interacting with item $i$: $\hat{y}_{ui}=\widetilde{\mathbf{e}}_u \cdot\widetilde{\mathbf{e}}_i$. 
    
\subsection{Synergistic Contrastive Learning (SCL). } 
 The fused embeddings from the two perspectives—collaborative filtering (first-order) and hypergraph (second-order)—are further refined via a synergistic cross-view contrastive learning mechanism. 
 We generate two “views” of each user and item: one from the hypergraph side ($\mathbf{h}_{u}$ or $\mathbf{h}_{i}$) and one from the fused side, which includes the ID embedding ($\widetilde{\mathbf{e}}_u$ or $\widetilde{\mathbf{e}}_i$). 
 For each user $u$, we treat $(\mathbf{h}_{u}, \widetilde{\mathbf{e}}_u)$ as a positive pair since they correspond to the same user’s information. 
 Similarly, for each item $i$, $(\mathbf{h}_{i}, \widetilde{\mathbf{e}}_i)$ is a positive pair. 
 Pairs of representations from different users (or different items) are treated as negatives. 
 We then define contrastive loss functions for the user side and item side. 
 Specifically, we use a normalized temperature-scaled dot product as the similarity measure $s(\mathbf{x}_1,\mathbf{x}_2)=\mathbf{x}_1^\top\cdot\mathbf{x}_2/(\tau\cdot\|\mathbf{x}_1\|_2\cdot\|\mathbf{x}_2\|_2)$, where $\tau$ is a temperature hyperparameter. 
 The contrastive learning loss functions at both the user and item levels are as follows:  
 \begin{equation}
    \begin{split}
        \mathcal{L}_{\mathrm{u2u}}&=-\sum_{u\in\mathcal{U}}^{M}\log\frac{\exp s(\mathbf{h}_u,\widetilde{\mathbf{e}}_u)}{\sum_{u^{\prime}\in\mathcal{U}}\left(\exp s(\mathbf{h}_{u^{\prime}},\widetilde{\mathbf{e}}_u)+\exp s(\widetilde{\mathbf{e}}_{u^{\prime}},\widetilde{\mathbf{e}}_u)\right)} \\
        \mathcal{L}_{\mathrm{i2i}}&=-\sum_{i\in\mathcal{I}}^{N}\log\frac{\exp s(\mathbf{h}_i,\widetilde{\mathbf{e}}_i)}{\sum_{i^{\prime}\in\mathcal{I}}\left(\exp s(\mathbf{h}_{i^{\prime}},\widetilde{\mathbf{e}}_i)+\exp s(\widetilde{\mathbf{e}}_{i^{\prime}},\widetilde{\mathbf{e}}_i)\right)}. 
         \end{split}
    \end{equation}
 In essence, for each user (or item), we maximize the agreement between its hypergraph view and fused view, while minimizing the agreement with other users’ views. By doing so, we encourage the model to learn user and item representations that are consistent across the two perspectives. The contrastive terms push $\widetilde{\mathbf{e}}_u$ to be closer to $\mathbf{h_u}$ (and $\widetilde{\mathbf{e}}_i$ closer to $\mathbf{h_i}$) in the representation space, effectively transferring the high-order semantic information into the fused embeddings. This cross-view alignment provides an auxiliary self-supervised signal that helps to alleviate data sparsity: even if a user has very few interactions (sparse first-order signal), the hypergraph view $\mathbf{h_u}$ can still influence $\widetilde{\mathbf{e}}_u$ through $\mathcal{L}_{u2u}$, and similarly for items via $\mathcal{L}_{i2i}$. Overall, the contrastive learning component enhances feature robustness and distinguishability by leveraging the rich second-order information captured in the hypergraphs.

\subsection{Loss Function}
The final loss function of MMHCL combines the standard recommendation loss with the contrastive losses described above, along with regularization:
    \begin{equation}
        \mathcal{L}=\mathcal{L}_{\mathrm{BPR}}+\alpha\cdot\mathcal{L}_{\mathrm{u2u}}+\beta\cdot\mathcal{L}_{\mathrm{i2i}}+\lambda \cdot\|\Theta\|^{2},
    \end{equation}
    where $\alpha$ and $\beta$ represent weights assigned to $\mathcal{L}_{\mathrm{u2u}}$ and $\mathcal{L}_{\mathrm{i2i}}$, respectively, which serve to balance the contributions of the two hypergraphs synergistically. 
    The last term is the $L_{2}$ regularization with coefficient $\lambda$ to prevent overfitting.  
    We adopt the Bayesian Personalized Ranking (BPR) loss, a pairwise approach that predicts higher scores for observed entries compared to unobserved ones: 
    \begin{equation}
        \mathcal{L}_{\mathrm{BPR}}=-\sum_{u=1}^{M}\sum_{i\in\mathcal{N}_{u}}\sum_{j\notin\mathcal{N}_{u}}\ln\sigma(\hat{y}_{ui}-\hat{y}_{uj}),
    \end{equation}
    where $\mathcal{N}_{u}$ denotes the set of items that interact with the user. 
    $i\in\mathcal{N}_{u}$ is the positive item and $j\notin\mathcal{N}_{u}$ is the negative item sampled from unobserved interactions. 
    $\sigma(\cdot)$ refers to the sigmoid function. 

\section{EXPERIMENTS}

\subsection{Experimental Settings}
\subsubsection{Datasets}

    To comprehensively evaluate our proposed method, we conduct extensive experiments on three widely used and publicly available datasets: TikTok, Clothing, and Sports. 
    The details of datasets are shown in Table \ref{Table:1}. 
    The Tiktok dataset, sourced from the TikTok platform to record users' interactions with short videos they've viewed \cite{wei2023multi}, includes visual, acoustic, and textual modalities. 
    (a) Clothing, Shoes and Jewelry (Clothing) and (b) Sports and Outdoors (Sports) datasets are two benchmark datasets from Amazon \cite{mcauley2015image}. 
    These two datasets include visual and textual modalities. 
    We use the same datasets and feature processing methods as in previous works ~\cite{wei2023multi,zhang2021mining}.  
    Additionally, we utilize pre-extracted and publicly available multimodal features in our experiments. 
    By strictly following the employed protocol in previous literatures ~\cite{wei2023multi,zhang2021mining}, we randomly divide each dataset into training, validation, and test sets in an 8:1:1 ratio. 
\begin{table}[h]
 
    \renewcommand\arraystretch{1.3}
    \caption{Statistics of the datasets with multimodal item Visual(V), Acoustic(A), Textual(T) contents.}
    \begin{center}
    \setlength{\tabcolsep}{0.5mm}
        \begin{tabular}{ccccccccccc}
        \hline
            Dataset &   &   \multicolumn{3}{c}{Tiktok}  &   &   \multicolumn{2}{c}{Clothing}    &   &   \multicolumn{2}{c}{Sports}  \\ \hline 
            Modality  &    & V & A & T & & V & T &  & V & T \\   
            Embed Dim &    & 128 & 128 & 768 & & 4096 & 1024  & & 4096 & 1024      \\ \cline{1-1} \cline{3-5} \cline{7-8} \cline{10-11}  
            User  &    & \multicolumn{3}{c}{9319} & & \multicolumn{2}{c}{39387}  &  & \multicolumn{2}{c}{35598}   \\ 
            Item   &    & \multicolumn{3}{c}{6710} & & \multicolumn{2}{c}{23033}  &  & \multicolumn{2}{c}{18357}    \\  
            Interactions   &    & \multicolumn{3}{c}{59541} & & \multicolumn{2}{c}{237488}  &  & \multicolumn{2}{c}{256308}   \\ \hline 
            Sparsity   &    & \multicolumn{3}{c}{99.904\%} & & \multicolumn{2}{c}{99.974\%}  &  & \multicolumn{2}{c}{99.961\%}   \\ \hline 
        \end{tabular}
    \end{center}
   \label{Table:1}
\end{table}

\subsubsection{Evaluation Metrics}

    For a fair comparison, we employ three widely used metrics in the recommendation to evaluate the effectiveness of preference ordering: Recall@K (R@K), Precision@K (P@K), and Normalized Discounted Cumulative Gain (N@K). 
    We set K to 20, consistent with previous studies  ~\cite{wei2023multi}, and report the average metrics computed across all users in the test set. 
    

\subsubsection{Baseline Methods} 

    To validate the effectiveness of our model, we compare it with various state-of-the-art (SOTA) models, including two graph-based collaborative filtering models (NGCF \cite{wang2019neural}, LightGCN \cite{he2020lightgcn}), recent self-supervised recommendation approaches (SGL \cite{wu2021self} ), as well as seven multimodal-based recommendation methods (VBPR \cite{he2016vbpr}, MMGCN \cite{wei2019mmgcn}, GRCN \cite{wei2020graph}, LATTICE \cite{zhang2021mining}, SLMRec \cite{tao2022self}, FREEDOM \cite{zhou2023tale}, MMSSL \cite{wei2023multi}). 
        \begin{itemize}
        \item \textbf{MF-BPR} \cite{rendle2012bpr}  optimizes the factorization process using the Bayesian Personalized Ranking (BPR) loss to enhance personalized ranking. 
	\item \textbf{NGCF} \cite{wang2019neural} represents user-item interactions as a bipartite graph and captures higher-order collaborative filtering signals using graph convolution. 
	\item \textbf{LightGCN} \cite{he2020lightgcn}  removes feature transformation and nonlinear activation in graph convolution. 
	\item \textbf{SGL} \cite{wu2021self} uses SSL for data augmentation through perturbation graph structure. 
	\item \textbf{VBPR} \cite{he2016vbpr} is the first work to combine multimodal features in recommendation. It incorporates visual features into the matrix decomposition. 
	\item \textbf{MMGCN} \cite{wei2019mmgcn} conducts graph convolution within each modality to capture diverse modal preferences of users. 
        \item \textbf{GRCN} \cite{wei2020graph} identifies false-positive feedback in graph convolution and cuts noisy edges. 
        \item \textbf{LATTICE} \cite{zhang2021mining} employs deep neural networks to mine latent structural information among items. 
        \item \textbf{SLMRec} \cite{tao2022self} generates various representation views by utilizing dropout and mask operations at the feature level for data augmentation. 
        \item \textbf{LGMRec} \cite{guo2024lgmrec} jointly models local and global user interests via local graph and global hypergraph embeddings to mitigate embedding coupling and interaction sparsity. 
        \item \textbf{FREEDOM} \cite{zhou2023tale} freezes the graph structure of LATTICE and removes redundant noise from the graph structure. 
        \item \textbf{MMSSL} \cite{wei2023multi} is a state-of-the-art multimodal recommendation method. It generates modality-aware interactive structures through adversarial perturbations and applies contrastive learning across modalities. 

    \end{itemize}

\subsubsection{Hyper-parameters Settings}
    

We implement our model in PyTorch and run all experiments on a single NVIDIA 3090 GPU.  To ensure fair comparison with prior work, we set the embedding dimension $d=64$ for all models (including baselines). Model parameters are initialized with Xavier initialization \cite{glorot2010understanding}. We train using the Adam optimizer \cite{kingma2014adam} with a learning rate of 0.0001 and a batch size of 1024. 
For our hypergraph construction, we search the number of nearest neighbors $K$ for the hypergraph from $\{5, 10, 15, 20, 25, 30\}$. 
The $\alpha$ and $\beta$ searched from $\{0.1, 0.3, 0.5, 0.7, 0.9\}$. 
The contrastive learning temperature $\tau$ is adjusted from 0.1 to 1 in 0.1 increments. 
The hyperparameter $\lambda$ for $L_{2}$ regularization is searched from  $\{$1e-1, 1e-2, 1e-3, 1e-4, 1e-5$\}$. 
    The depth of u2u hypergraph layers and i2i hypergraph layers searched from {1, 2, 3} with collaborative effect. 
    To ensure sufficient model convergence, the number of epochs is set to 250. Additionally, an early stopping criterion of 5 is applied to terminate training when no further performance improvement is observed on the validation set.
Table \ref{tab:hyperparameter_values} summarizes the key hyperparameters used for MMHCL on the three datasets.     For a detailed analysis of parameter experiments, see Section \ref{hyperparameters_detail}.

\begin{table}[htbp]
\centering
\caption{Key hyperparameter values for MMHCL on Tiktok, Clothing, and Sports datasets.}
\begin{tabular}{lccc}
\toprule
\textbf{Hyperparameter} & \textbf{Tiktok}   &   \textbf{Clothing} & \textbf{Sports} \\
\midrule
Embedding dimension ($d$) & 64 & 64 & 64 \\
Learning rate & 0.0001 & 0.0001 & 0.0001\\
Batch size & 1024 & 1024 & 1024\\

Number of nearest neighbors $K$ for hypergraph  &   5   &   10  &   5 \\
$\alpha$ for u2u hypergraph &   0.03    &   0.1 &   0.3 \\
$\beta$ for i2i hypergraph  &   0.07    &   0.7 &   0.7 \\
Contrastive learning temperature $\tau$ &   0.6 &   0.4 &   0.5 \\
$L_{2}$ regularization coefficient $\lambda$    &   1e-3    &   1e-3    &   1e-5 \\
u2u hypergraph layers   &   3   &   2   &   2 \\
i2i hypergraph layers   &   2   &   2   &   2  \\

Epoch & 250 & 250 & 250 \\
Early stopping & 5 & 5 & 5\\
\bottomrule
\end{tabular}
\label{tab:hyperparameter_values}
\end{table}

\subsection{Performance Comparison}
\subsubsection{Overall Performance}
\begin{table*}[tp]
        \caption{Performances of all comparison methods on three datasets. The best results are highlighted in bold, while the second-best ones are underlined. All improvements are statistically significant with a $p$-value $\leq 0.01$. }
	\label{Table:2}
	\renewcommand\arraystretch{1.2}
	\begin{center}
		{
		\begin{threeparttable}{
			\begin{tabular}{*{10}{c}}
				\toprule
				\multirow{2}{*}{Baseline} &
				\multicolumn{3}{c}{Tiktok} & \multicolumn{3}{c}{Clothing} & \multicolumn{3}{c}{Sports} \cr
				\cmidrule(lr){2-4}\cmidrule(lr){5-7}\cmidrule(lr){8-10} & R@20 & P@20 & N@20 & R@20 & P@20 & N@20 & R@20 & P@20 & N@20 \\ \hline
                    MF-BPR	& 0.0345 & 0.0018 & 0.0131 & 0.0191 & 0.0010 & 0.0088 & 0.0431 & 0.0024 & 0.0203 \\
                    NGCF	& 0.0603 & 0.0031 & 0.0239 & 0.0387 & 0.0020 & 0.0168 & 0.0696 & 0.0036 & 0.0319 \\
                    LightGCN & 0.0654 & 0.0032 & 0.0281 & 0.0470 & 0.0024 & 0.0215 & 0.0781 & 0.0041 & 0.0370 \\ 
                    SGL & 0.0603 & 0.0030 & 0.0238 & 0.0598 & 0.0030 & 0.0268 & 0.0779 & 0.0041 & 0.0361 \\ \hline
                    VBPR & 0.0380 & 0.0018 & 0.0134 & 0.0481 & 0.0024 & 0.0205 & 0.0582 & 0.0031 & 0.0265 \\
                    MMGCN & 0.0729 & 0.0037 & 0.0306 & 0.0501 & 0.0024 & 0.0221 & 0.0639 & 0.0033 & 0.0278 \\
                    GRCN & 0.0805 & 0.0037 & 0.0349 & 0.0631 & 0.0032 & 0.0276 & 0.0834 & 0.0045 & 0.0378 \\
                    LATTICE & 0.0843 & 0.0042 & 0.0367 & 0.0710 & 0.0036 & 0.0316 & 0.0915 & 0.0048 & 0.0424 \\
                    SLMRec & 0.0845 & 0.0042 & 0.0353 & 0.0670 & 0.0035 & 0.0299 & 0.0829 & 0.0043 & 0.0376 \\
                    LGMRec  &   0.0709  &   0.0035  &    0.0279 &   0.0781  &   0.0040  &   0.0345  &   \underline{0.1007}  &   \underline{0.0053}  &   0.0451 \\
                    FREEDOM &  0.0859 & 0.0043 & 0.0392 & \underline{0.0812}  & \underline{0.0041} & \underline{0.0359} & 0.0987 & 0.0052 & 0.0436 \\
                    MMSSL & \underline{0.0921} & \underline{0.0046} & \underline{0.0392} & 0.0740 & 0.0037 &0.0330 & 0.0998 & 0.0052 & \underline{0.0470} \\ \hline
                    MMHCL & \textbf{0.1139} & \textbf{0.0057} & \textbf{0.0492} & \textbf{0.0881} & \textbf{0.0045} & \textbf{0.0394} & \textbf{0.1064} & \textbf{0.0056} & \textbf{0.0501} \\ 
                    Improv. & 23.67\% & 23.91\% & 25.51\% & 8.50\% & 9.76\% & 9.75\% & 5.66\% & 5.66\% & 6.60\% \\ \hline
		      \end{tabular}}  
		\end{threeparttable}}
	\end{center}
        \end{table*}
    The experimental results are provided in Table. \ref{Table:2}, from which we can observe: 
    \begin{itemize}
    \item \textbf{Superior Performance of MMHCL:} 
    Our proposed MMHCL model consistently achieves state-of-the-art results across all datasets and metrics, underscoring the robustness and efficacy of our approach. 
    In particular, MMHCL significantly outperforms the strongest baseline models in terms of Recall@20, with improvements of 23.67\%, 8.50\%, and 5.66\% on the Tiktok, Clothing, and Sports datasets, respectively. 
    This highlights MMHCL's ability to effectively capture shared user preferences and intricate multimodal semantic relationships among items. 
    The utilization of detailed hypergraphs facilitates the learning of dense attribute representations, while the synergistic contrastive learning mechanism introduces auxiliary supervisory signals that enhance downstream task performance. 
    \item \textbf{Impact of Graph Neural Networks:} 
    When compared to matrix decomposition-based collaborative filtering methods (e.g., MF-BPR, VBPR), GNN-based approaches such as LightGCN and MMGCN exhibit marked improvements in performance. 
    This result demonstrates the strength of GNNs in capturing multi-hop neighborhood information and extracting collaborative filtering signals via the message-passing mechanism. 
    These findings emphasize the importance of leveraging GNN architectures to enrich user-item representations and amplify interaction signals. 
    \item \textbf{Advantage of Multimodal Information:} 
    Models like MMGCN, which integrate multimodal features, show substantial gains over traditional collaborative filtering methods (e.g., NGCF, LightGCN). 
    Including multimodal information significantly improves user and item representation learning, indicating the necessity of incorporating these additional data modalities. 
    Such results emphasize the essential of multimodal data in enhancing the performance of recommendation systems by offering richer contextual information. 
    \item \textbf{Benefits of Self-Supervised Learning:}
    Recent methods that incorporate self-supervised learning (e.g., SLMRec, MMSSL) generally outperform others, indicating their ability to mitigate the data sparsity problem through auxiliary supervisory signals. 
    Compared with other multimodal recommendation methods that incorporate self-supervised learning (e.g., SLMRec, MMSSL), our approach has a significant performance improvement. 
    Our analysis suggests that previous SSL-based methods (e.g., SLMRec) utilizing masking or dropout operations at the feature or modality levels may inadvertently discard valuable information leading to performance degradation. 
    The approaches (e.g. MMSSL) focusing on user-item interactions may not effectively utilize correlations in items' multimodal information due to data sparsity. 
    \item \textbf{MMHCL's Distinctive Mechanism:}
     Unlike prior SSL-based methods, our MMHCL introduces a unique approach that simultaneously considers both user and item levels by utilizing hypergraphs to uncover richer second-order semantic information. 
     The item-to-item hypergraph reduces reliance on explicit user-item interactions, which is particularly advantageous in sparsely populated datasets. 
     Additionally, the contrastive feature enhancement paradigm we propose ensures that contrastive learning occurs at both user and item levels. 
     This mechanism prevents the loss of critical information and further alleviates the data sparsity issue by providing more robust auxiliary supervisory signals. 
    \end{itemize}
    In summary, the experimental results clearly demonstrate the efficacy of our MMHCL model across multiple dimensions. 
    The integration of hypergraph learning with multimodal contrastive signals presents a robust framework for handling sparsity and maximizing the utilization of multimodal data, setting a new benchmark for state-of-the-art performance in multimodal recommendation.
\begin{figure}[t]
	\centering
 \setlength{\abovecaptionskip}{0.2cm}
	\includegraphics[width=0.5\textwidth]{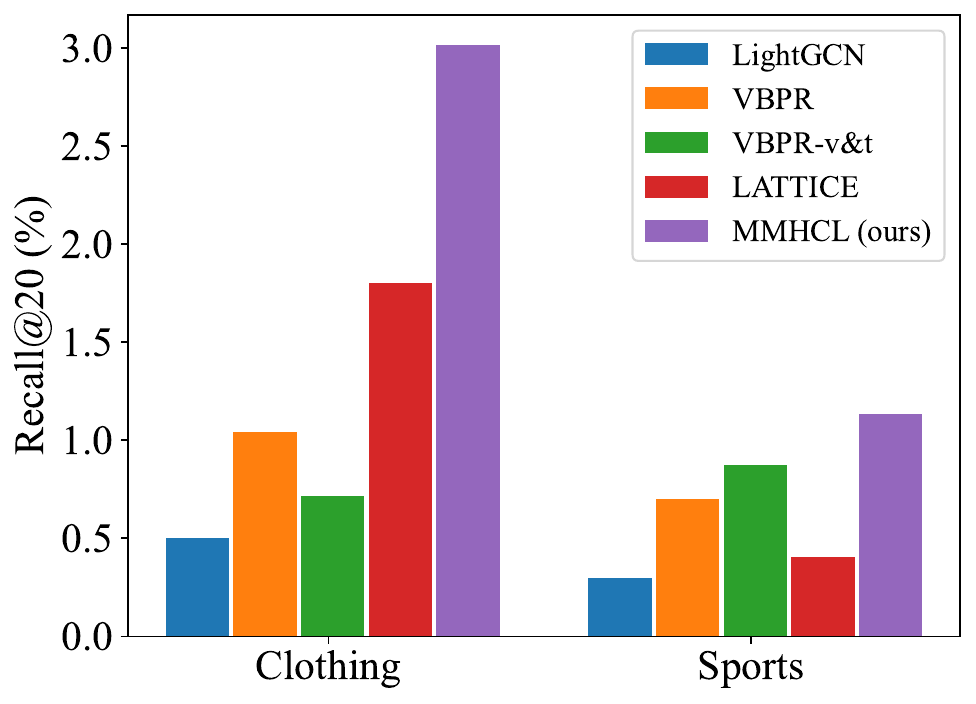}
	\caption{Performance in cold-start problem.}
	\label{cold}
	\vspace{-5pt}
\end{figure}

\subsubsection{Performance in cold-start problem}


   As shown in Fig. \ref{cold}, we conducted cold-start experiments on the Clothing and Sports datasets to further assess the robustness of our model, following the methodology outlined in prior work \cite{zhang2021mining}. Specifically, we randomly selected 20\% of the item nodes from the training set and removed all user associations related to these items, effectively excluding them from the training phase. This simulates a cold-start scenario where no prior interaction data exists for these items.

    The results demonstrate that our proposed MMHCL model consistently outperforms all baseline methods, including LightGCN, VBPR, VBPR-v\&t, and LATTICE, across both datasets. Notably, all multimodal-based approaches outperformed LightGCN, highlighting the critical importance of integrating multimodal information into recommendation systems. LightGCN, which relies solely on user-item interaction data during training, suffers from a significant performance drop in cold-start scenarios. The absence of interaction information prevents the cold-start item representations from being effectively updated through backpropagation, limiting the model’s ability to generalize.

    In contrast, our approach demonstrates superior cold-start performance by leveraging rich semantic information from both user and item perspectives, independent of direct user-item interactions. By constructing hypergraphs that capture second-order relationships and multimodal associations, MMHCL effectively mitigates the cold-start issue. The hypergraph structure allows for a more robust and nuanced fusion of multimodal data, enabling the model to generate more comprehensive item representations even in the absence of interaction history.

    Overall, these results underscore the effectiveness of our method in addressing the cold-start problem and the advantages of using hypergraphs to fuse multimodal information. Our approach not only alleviates the limitations of traditional collaborative filtering methods but also provides a scalable solution for improving recommendation accuracy in scenarios with sparse user-item interactions.

\subsection{Ablation Study}
\begin{table}[t]
    \caption{Performance comparison between MMHCL and its variants.}
    \centering
    \begin{tabular}{c|cc|cc|cc}
        \hline
        Data & \multicolumn{2}{c|}{Tiktok} & \multicolumn{2}{c|}{Clothing} & \multicolumn{2}{c}{Sports}\\
        \hline
        Metrics & R@20 & N@20 & R@20 & N@20 & R@20 & N@20\\
        \hline
        w/o-u2u & 0.1031 & 0.0459 & 0.0753 & 0.0337  & 0.0870 & 0.0396 \\
        w/o-i2i  & 0.1021 & 0.0432 &  0.0587 & 0.0271 & 0.0949 & 0.0451 \\
        w/o-scl  &  0.0997 & 0.0457 &  0.0744 & 0.0332 & 0.0971 & 0.0443 \\
        \hline
         \textbf{MMHCL} & \textbf{0.1139} & \textbf{0.0492} & \textbf{0.0881} & \textbf{0.0394} & \textbf{0.1064} & \textbf{0.0501} \\
        \hline
    \end{tabular}
    \label{Table:3}
    \vspace{-5pt}
\end{table}

To evaluate the effectiveness of our proposed components and the sensitivity of hyper-parameters, we conduct comprehensive ablation studies.  We design three model variants: (i) w/o-u2u: removing the user-to-user hypergraph (i.e., no $\mathbf{h}_u$; users are only represented by LightGCN’s embeddings and any item hypergraph information); (ii) w/o-i2i: removing the item-to-item hypergraph (no $\mathbf{h}_i$); (iii) w/o-scl: removing the self-supervised contrastive learning module (i.e., setting $\alpha = \beta = 0$, but still performing the fusion of $\mathbf{h}_u$ and $\mathbf{h}_i$ into $\mathbf{e}_u$ and $\mathbf{e}_i$ as described in Eq.\ref{eq9} and Eq.\ref{eq10}).

The results, as illustrated in Table. \ref{Table:3}, highlight the critical importance of each component. When the i2i hypergraph is removed (w/o-i2i), a significant drop in performance is observed, particularly on the TikTok and Clothing datasets. This reduction can be attributed to the rich multimodal content of these datasets, where item-specific information plays a pivotal role in learning robust item and user representations. In these cases, the i2i hypergraph contributes substantially to capturing the intricate relationships between items from different modalities, which in turn enhances recommendation quality.

Conversely, on the Sports dataset, the absence of the u2u hypergraph (w/o-u2u) leads to the most significant performance degradation. This highlights the importance of shared interest preferences between users in this domain, where user-user relationships are more crucial than item-based similarities. The u2u hypergraph effectively captures these second-order user relationships, which are essential for improving the model’s performance in this context.

In all cases, excluding the contrastive learning module (w/o-scl) results in a noticeable decline across all datasets, reaffirming the importance of synergistic contrastive learning. By maximizing the mutual information between first-order and second-order embeddings, the contrastive learning mechanism enhances feature distinguishability, thus providing auxiliary supervised signals that alleviate data sparsity. The substantial performance decrease observed when this component is removed underscores its critical role in improving the robustness and generalization of the model.

Overall, the ablation results confirm that each component—whether the u2u hypergraph, the i2i hypergraph, or contrastive learning—plays a vital role in the MMHCL framework. The absence of any one of these components leads to a marked reduction in performance, validating the integral contribution of each module in effectively capturing complex higher-order semantic information and addressing the cold-start problem.

    \begin{figure}[tbp]
    \centering
    \includegraphics[width=0.45 \textwidth]{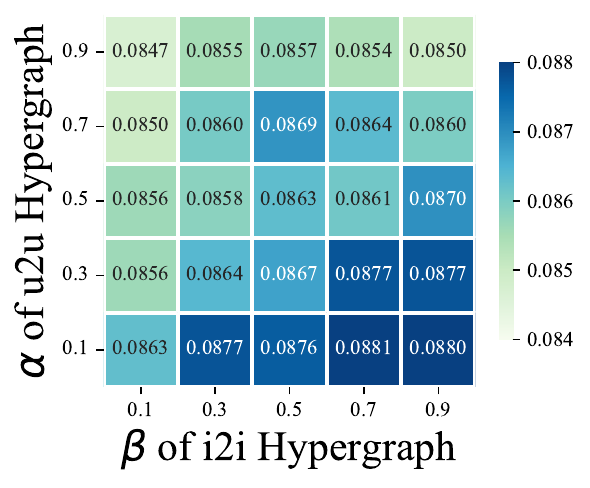}
    \includegraphics[width=0.45\textwidth]{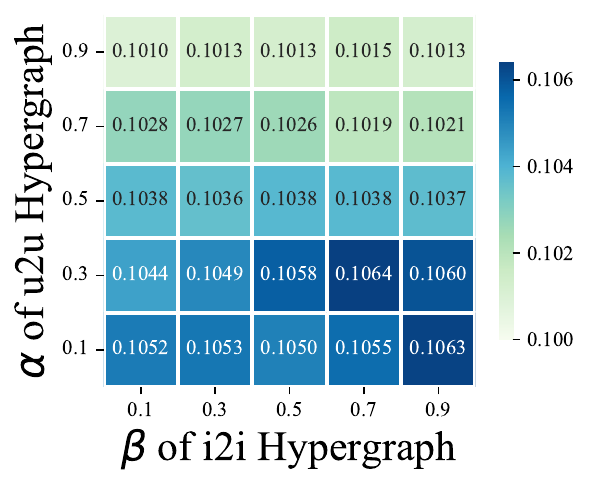}
    \caption{Performance w.r.t the effect of synergistic contrastive learning coefficients  $\alpha$ for u2u hypergraph and $\beta$ for i2i hypergraph on Clothing and Sports datasets (Recall@20). }
    \label{1.u2u_i2i_cl_ratio}
\end{figure}


\subsection{ Impact Study of Hyperparameters}
\label{hyperparameters_detail}
In this section, we conduct a comprehensive investigation into the effects of key hyperparameters on the performance of our MMHCL model. By analyzing the interaction between user-to-user (u2u) and item-to-item (i2i) hypergraphs, we aim to provide a deeper understanding of their respective roles in enhancing recommendation accuracy across multiple datasets.

\subsubsection{Impact of Synergistic Contrastive Learning}
\
    \begin{figure*}[tbp]
\centering   
\setcounter{subfigure}{0} 
\subfigure[$\text{top-}K$]{

\includegraphics[width=0.43\textwidth]{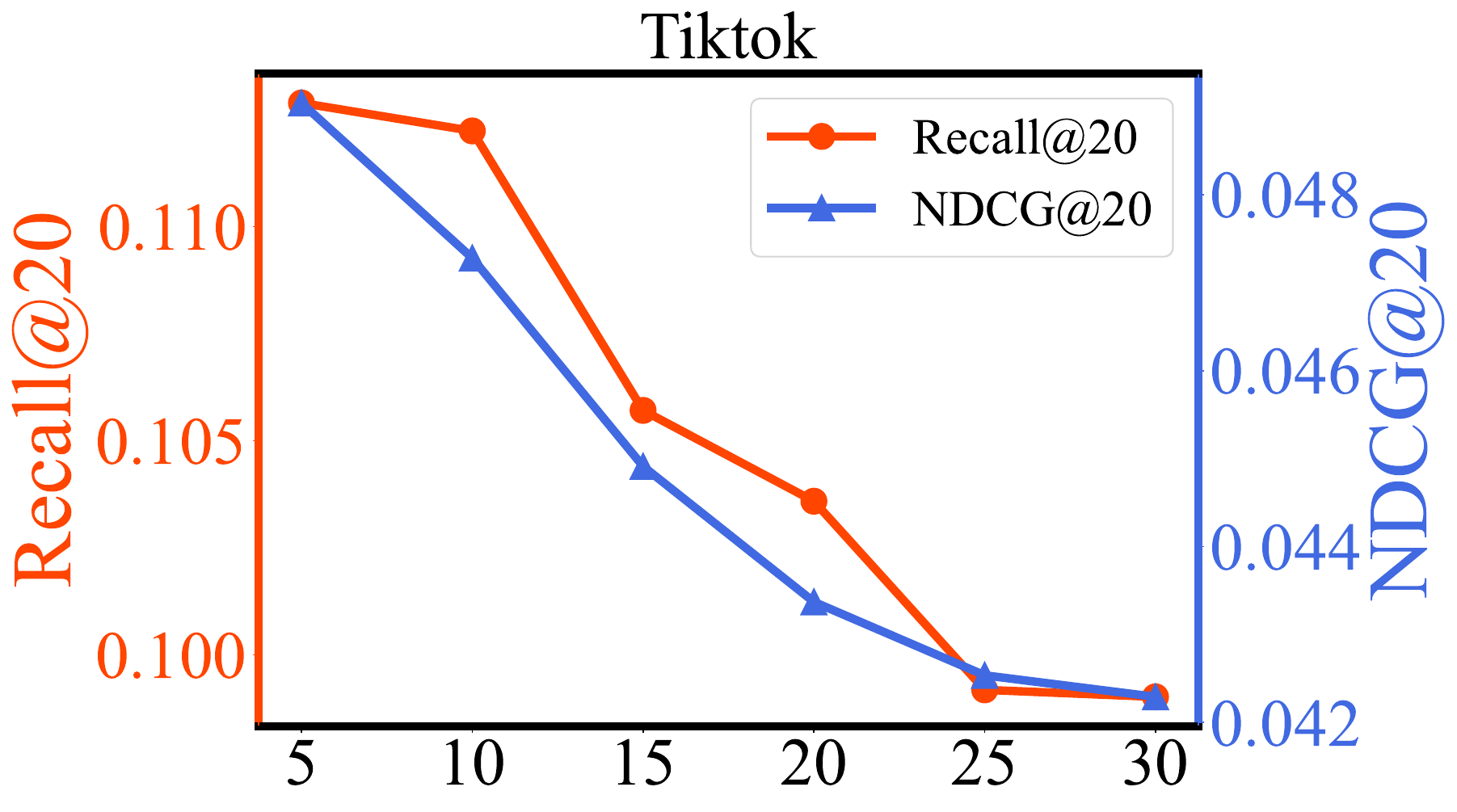}  

} 
\setcounter{subfigure}{3} 
\subfigure[Contrastive Learning Temperature $\tau$ ] {
\includegraphics[width=0.43\textwidth]{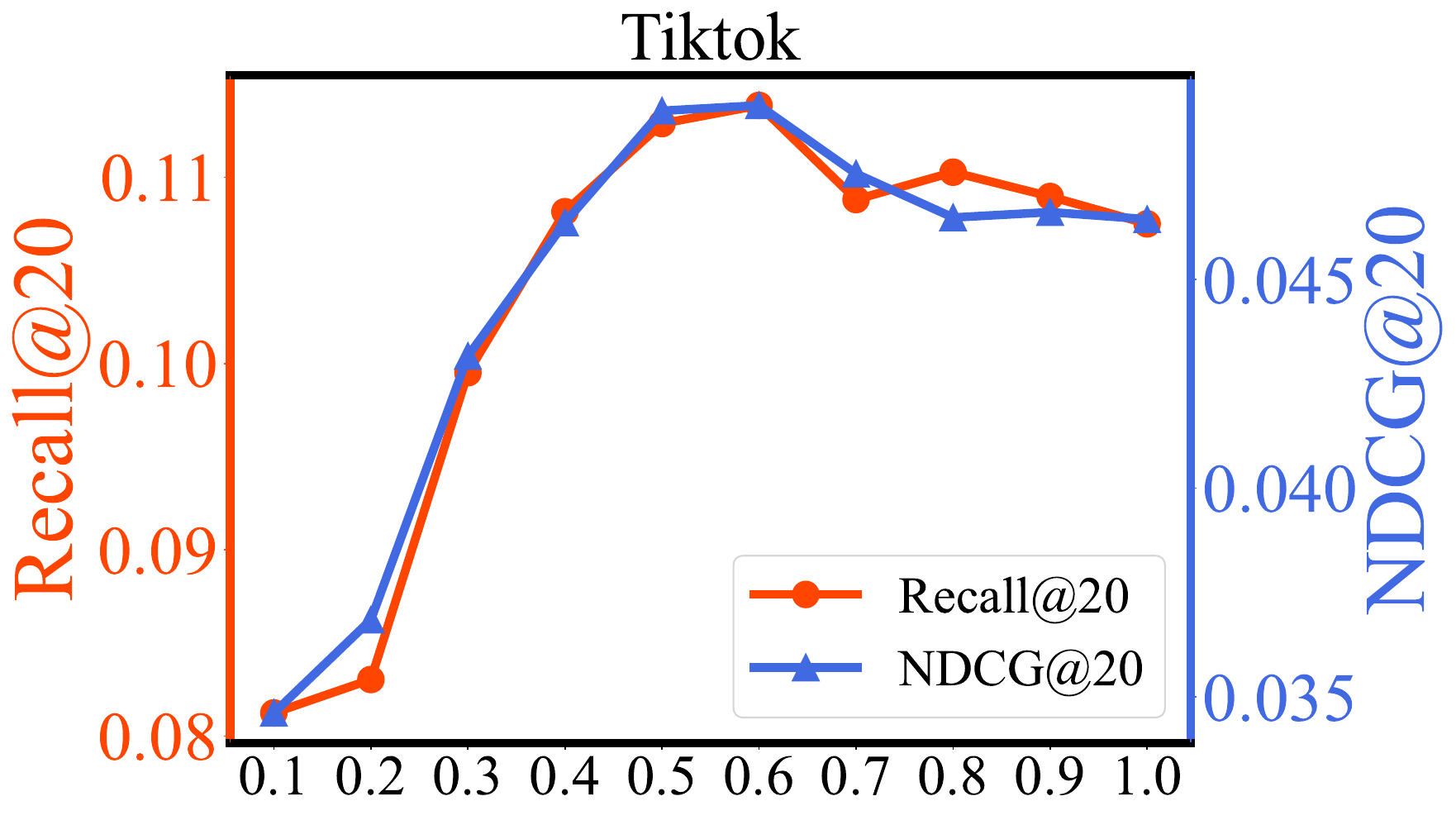}  
}     

\setcounter{subfigure}{1} 
\subfigure[$\text{top-}K$] {
\includegraphics[width=0.43\textwidth]{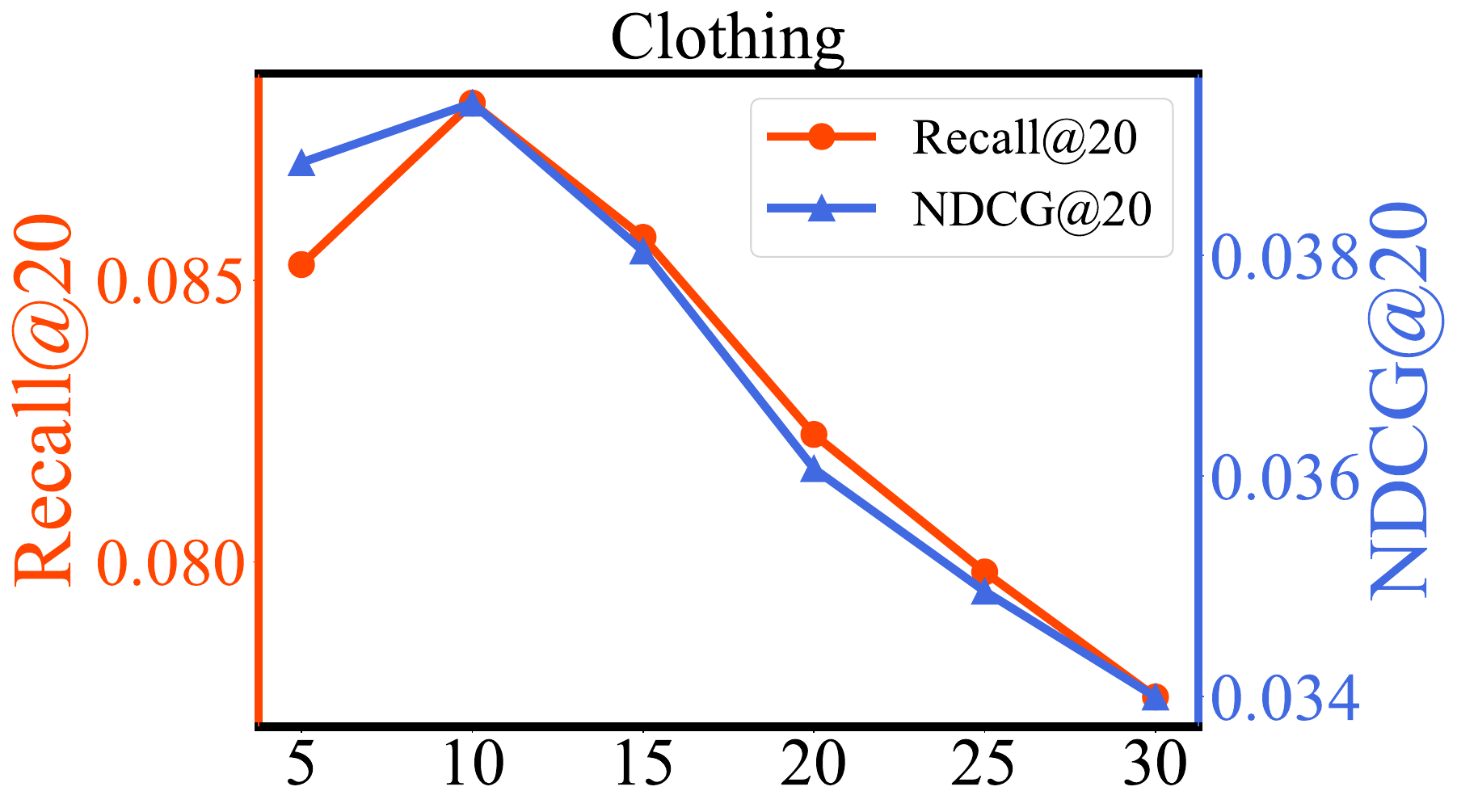}  
}     
\setcounter{subfigure}{4} 
\subfigure[Contrastive Learning Temperature $\tau$ ] {
\includegraphics[width=0.43\textwidth]{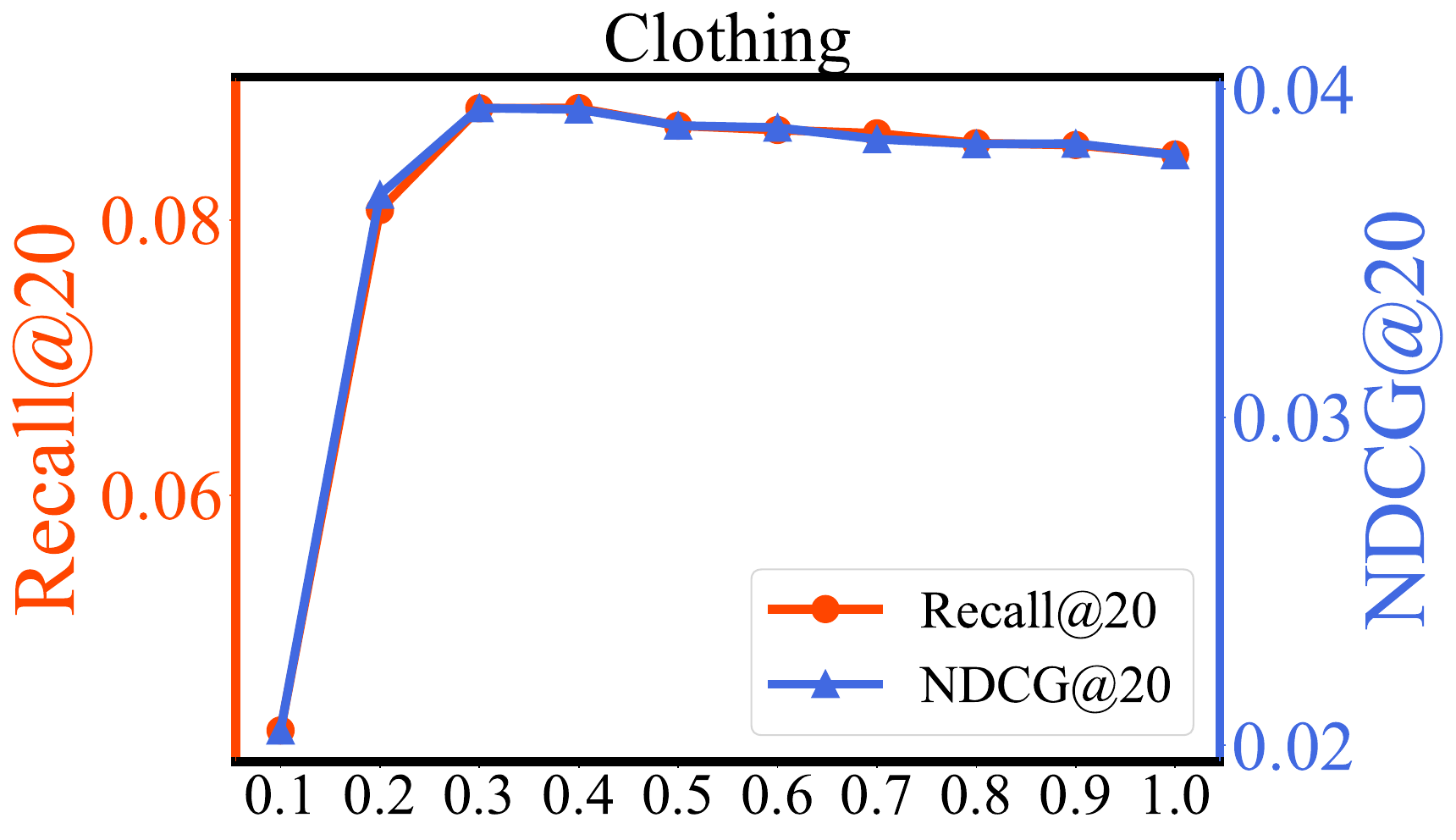}  
}     

\setcounter{subfigure}{2} 
\subfigure[$\text{top-}K$] {
\includegraphics[width=0.43\textwidth]{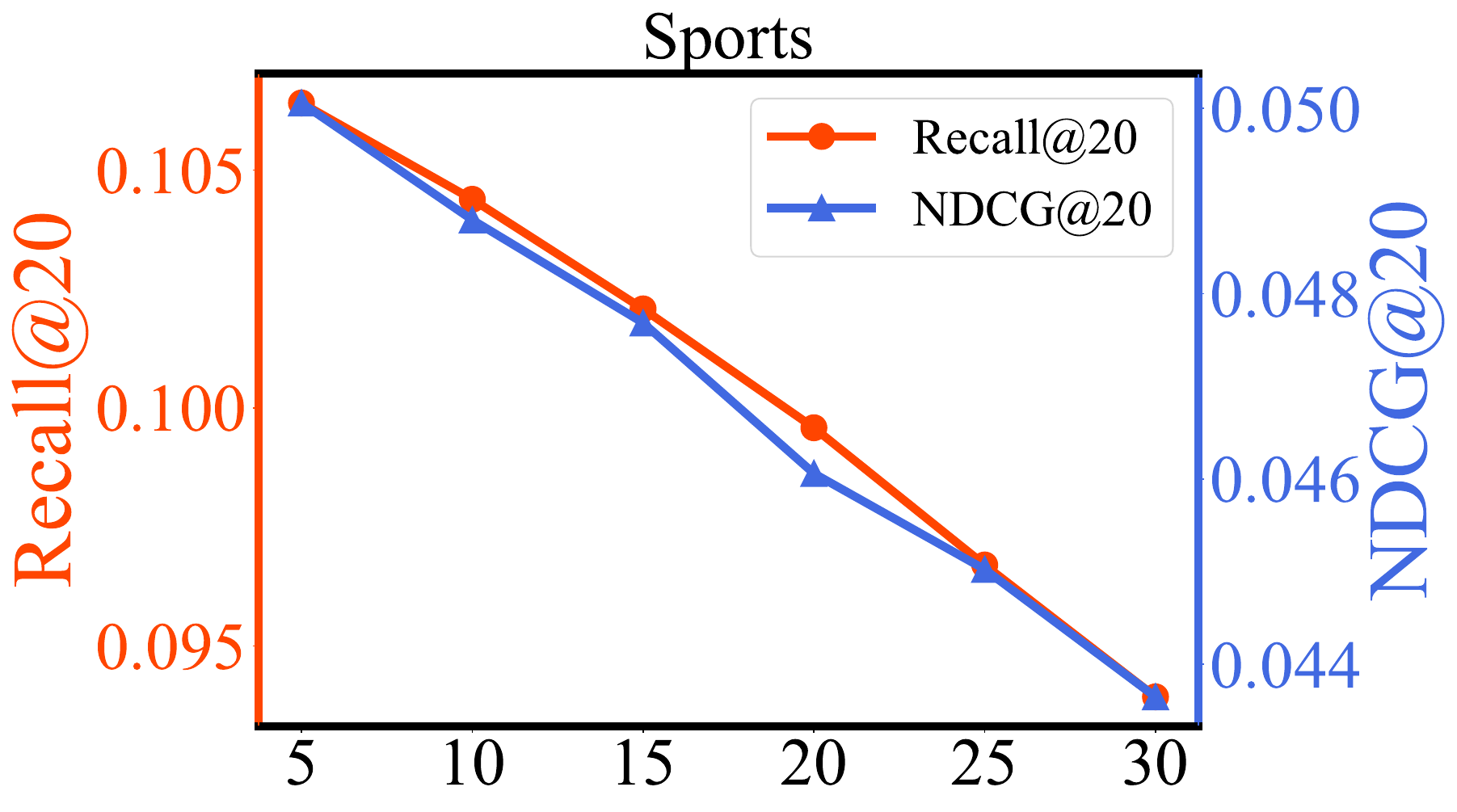}  
}     
\setcounter{subfigure}{5} 
\subfigure[Contrastive Learning Temperature $\tau$] {
\includegraphics[width=0.43\textwidth]{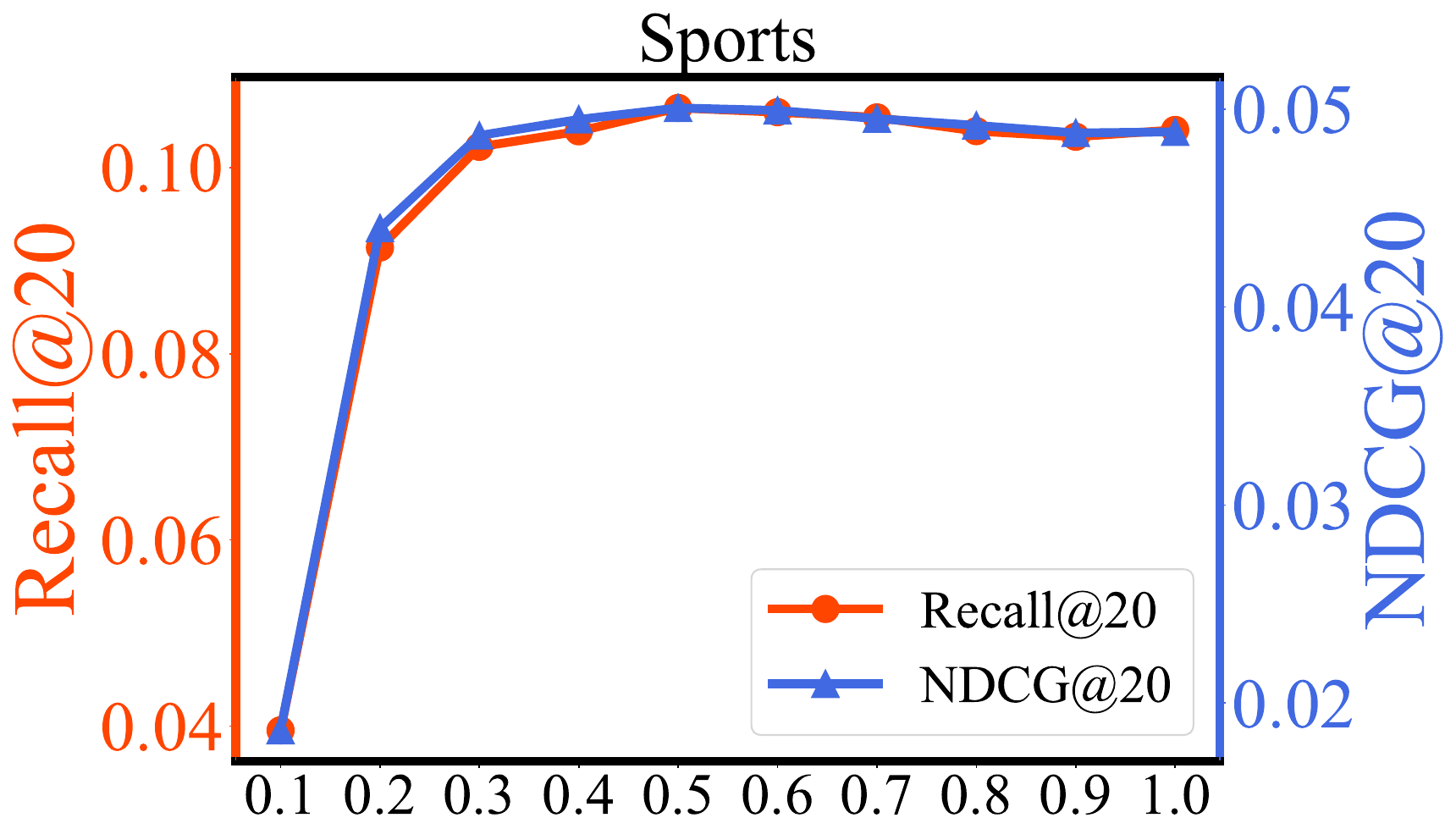}  
}     

\caption{Performance comparison w.r.t various hyperparameters on the three datasets.}
\label{hyperparameters1}
\end{figure*}

    We explored the synergistic effects of the u2u and i2i hypergraphs in our contrastive learning framework, and the results are visualized in Fig. \ref{1.u2u_i2i_cl_ratio} for the Clothing and Sports datasets. Consistent trends were observed in the TikTok dataset as well. In these experiments, $\alpha$ represents the contribution of the u2u hypergraph, and $\beta$ represents the contribution of the i2i hypergraph within the context of macro-level synergistic contrastive learning.

To maintain consistency, we set the number of convolutional layers for both hypergraphs to two. The results demonstrate a clear performance boost when the contribution of the i2i hypergraph ($\beta$) is increased, particularly when compared to configurations where $\alpha$ dominates. The results along the diagonal axis ($\alpha$ = $\beta$) suggest that a balanced incorporation of both hypergraphs yields robust performance. However, in most cases, the i2i hypergraph exhibits a stronger influence on the model’s effectiveness, as it effectively captures complex multimodal correlations between items, particularly in datasets rich with item-related information. 

These findings confirm that leveraging both hypergraphs in a synergistic manner, rather than in isolation, provides significant benefits. The macro-level contrastive learning framework effectively captures higher-order semantic relationships from both users and items, as demonstrated by the improved results over the w/o-u2u and w/o-i2i ablation experiments. This validates the importance of exploring contrastive learning from a dual perspective, encompassing both user and item levels, for improved recommendation performance.


The findings of our analysis highlight the pivotal role of contrastive learning in refining feature representations for both users and items by maximizing mutual information and amplifying feature discriminability. Specifically, the i2i hypergraph plays a crucial role in capturing intricate multimodal item relationships, facilitating more expressive item embeddings. Notably, this structured modeling of item interactions not only enhances representation granularity but also significantly improves the model’s generalization ability across diverse recommendation scenarios.

\newpage
\subsubsection{Impact of $\text{top-}K$ Values}
\


    The top-K value, which dictates the number of neighboring nodes considered during hypergraph convolution, plays a significant role in the MMHCL model’s performance. As illustrated in Fig. \ref{hyperparameters1}(a)-(c), smaller top-K values consistently lead to better performance across the TikTok, Clothing, and Sports datasets. This finding demonstrates that hypergraph convolution effectively aggregates relevant information from adjacent nodes at both user and item levels, resulting in superior model performance when fewer, more meaningful neighbors are considered.

    However, as top-K increases, the model’s performance degrades. This decline is due to the inclusion of irrelevant or noisy neighbors, which diminishes the quality of the aggregated information during hypergraph message passing. By introducing redundant connections, larger top-K values dilute the significance of the most relevant user-item relationships, ultimately reducing the precision of the model’s recommendations. These results underscore the need for careful sparsification of dense hypergraph structures to mitigate the adverse effects of noise and ensure the model focuses on the most informative connections.

\subsubsection{Impact of Contrastive Learning Temperature Varied $\tau$ }
\
    
    The contrastive learning temperature $\tau$ controls the strength of the contrastive loss by modulating the separation between positive and negative samples. As shown in Fig. \ref{hyperparameters1}(d)-(f), the model’s performance improves as $\tau$ increases initially, peaking at an optimal value before gradually declining as $\tau$ becomes too large. This trend is consistent across all three datasets.

When $\tau$ is too small, the model excessively separates positive samples from challenging negative samples, leading to the destruction of latent semantic relationships that are crucial for effective representation learning. In such cases, the contrastive learning mechanism becomes overly rigid, preventing the model from capturing subtle but important similarities between samples. On the other hand, when $\tau$ is too large, the model's ability to discriminate between samples diminishes, resulting in less effective feature representations. The observed performance drop at high $\tau$ values highlights the need for a balanced approach, where $\tau$ is carefully tuned to optimize the trade-off between separation and discrimination.

These results highlight the importance of selecting an appropriate contrastive learning temperature to maximize the model’s ability to learn meaningful user-item interactions while preserving the necessary flexibility to capture latent semantic structures.

\begin{figure*}[tbp]
\centering   

\setcounter{subfigure}{0} 
\subfigure[ $L_{2}$ Regularization Coefficient $\lambda$] {
\includegraphics[width=0.43\textwidth]{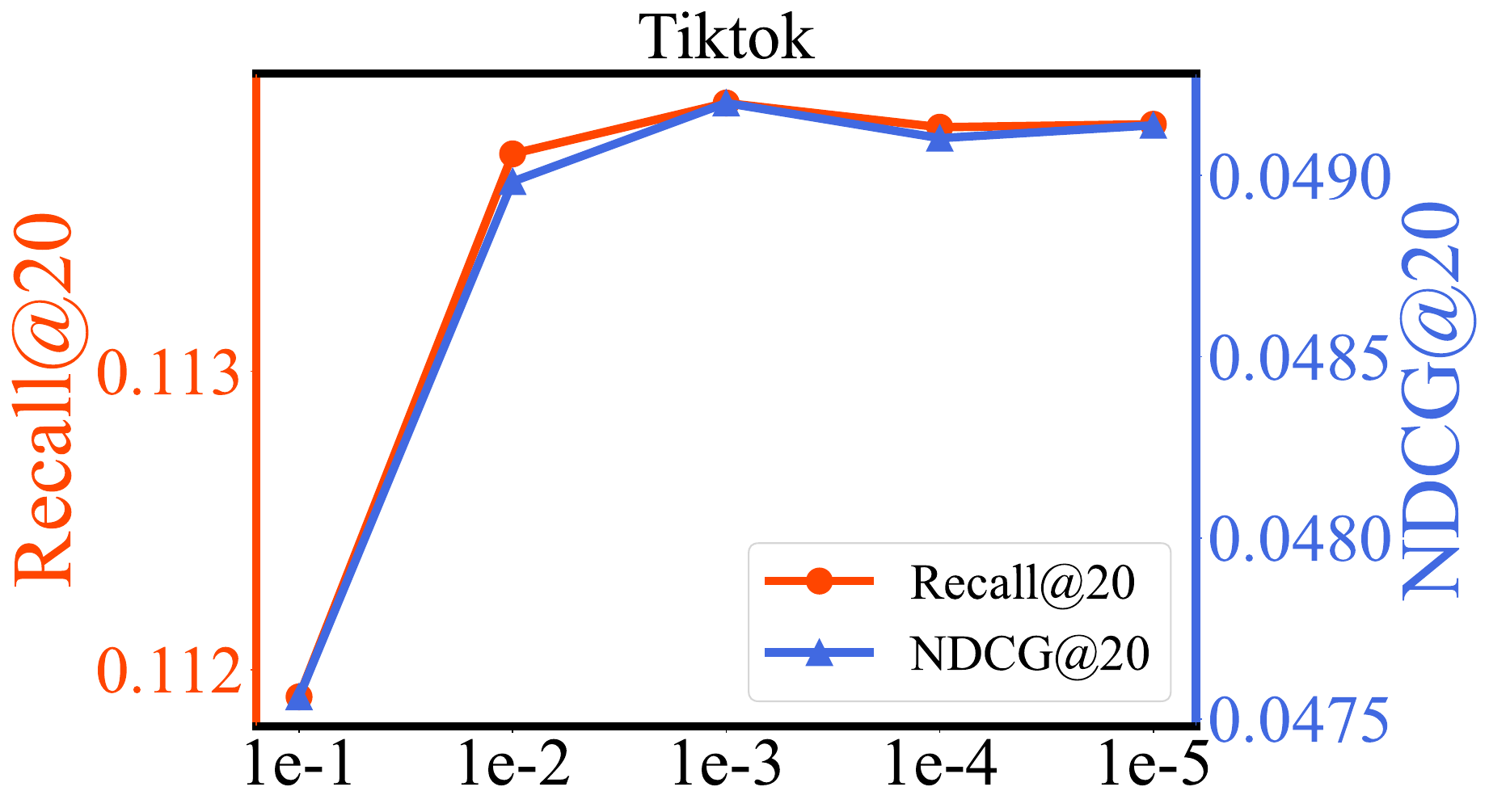}  
}     
\setcounter{subfigure}{3} 
\subfigure[dimension $d$] {
\includegraphics[width=0.43\textwidth]{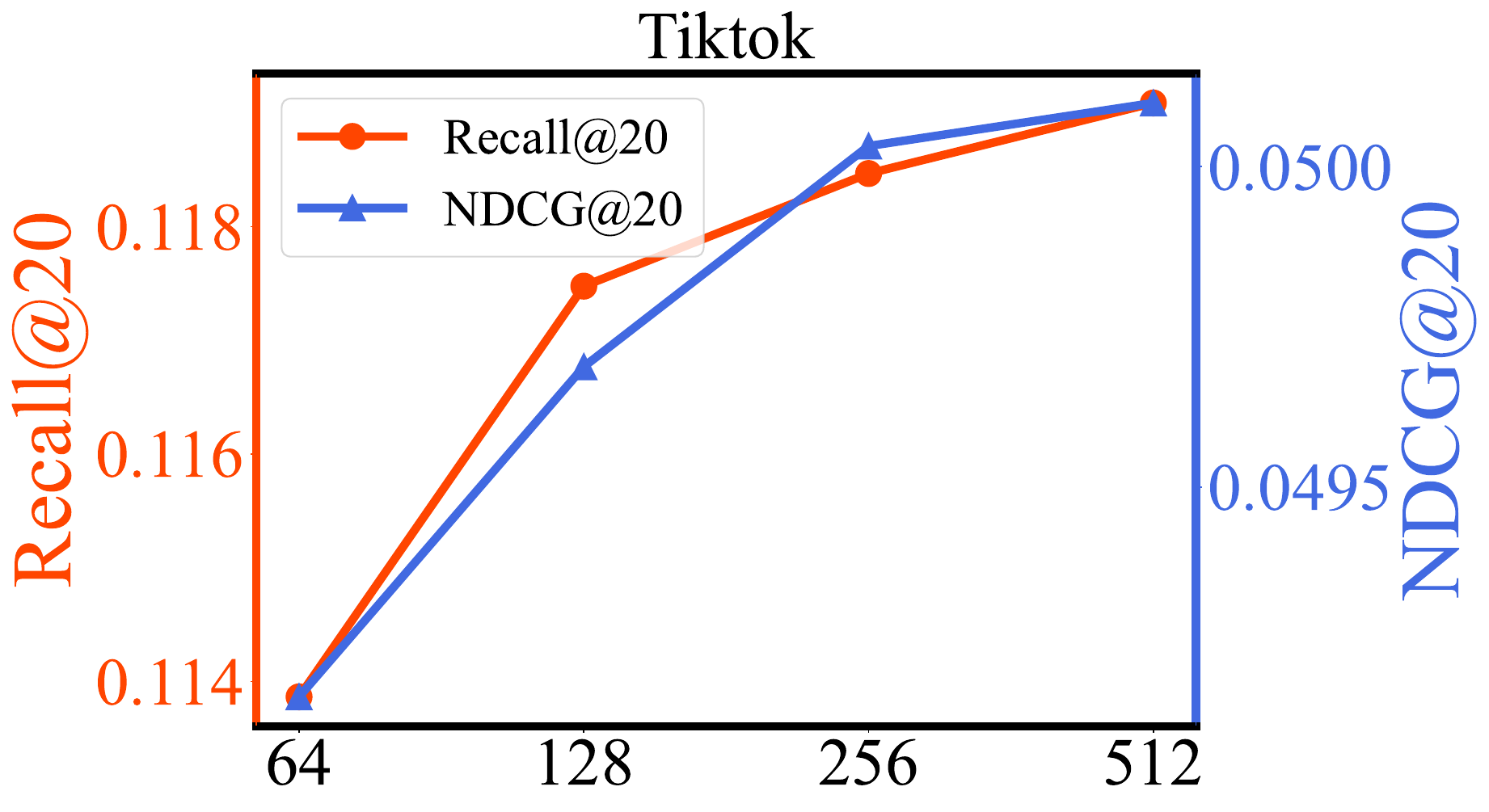}  
}     

\vspace{-3mm}

\setcounter{subfigure}{1} 
\subfigure[$L_{2}$ Regularization Coefficient $\lambda$] {
\includegraphics[width=0.43\textwidth]{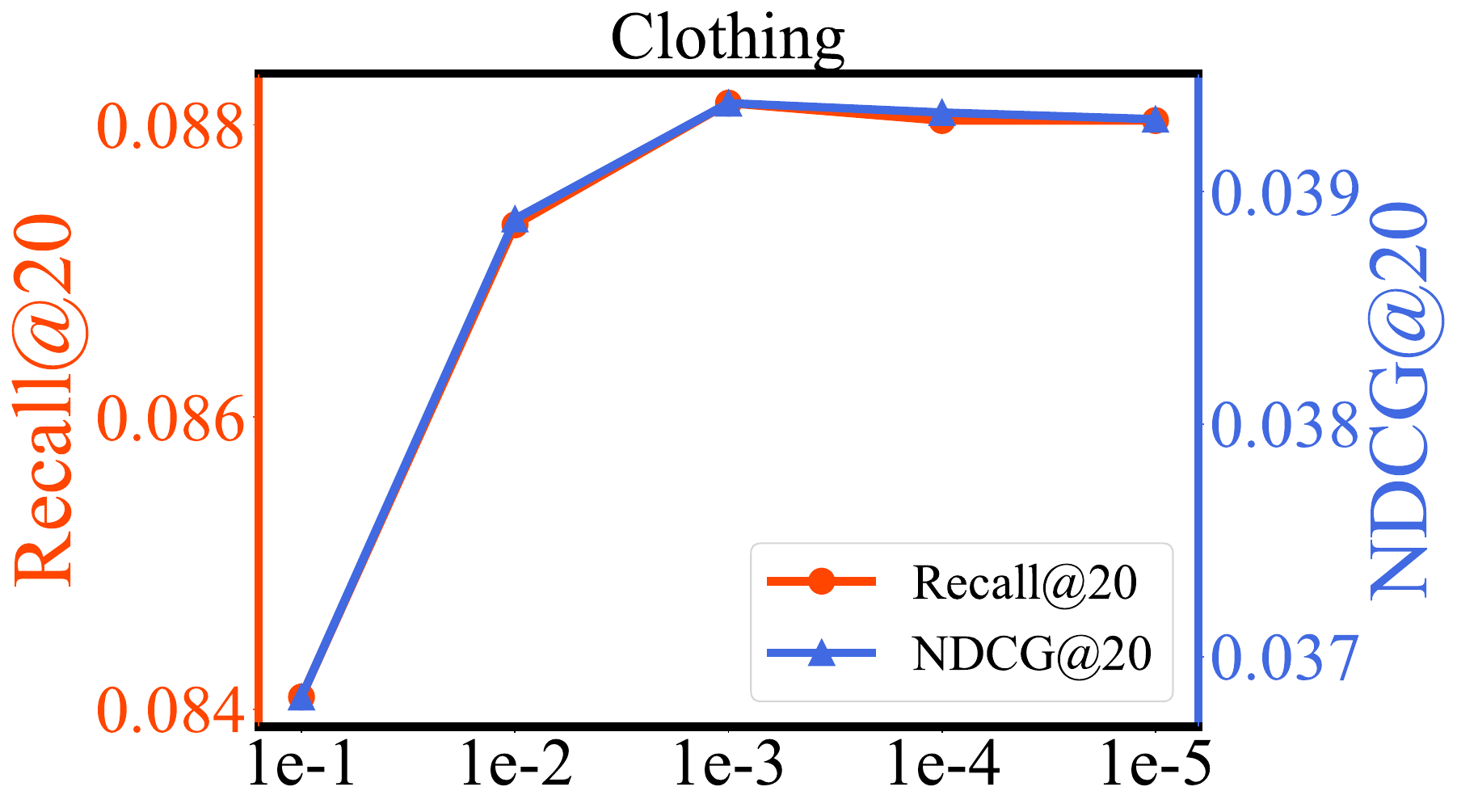}  
}     
\setcounter{subfigure}{4} 
\subfigure[dimension $d$] {
\includegraphics[width=0.43\textwidth]{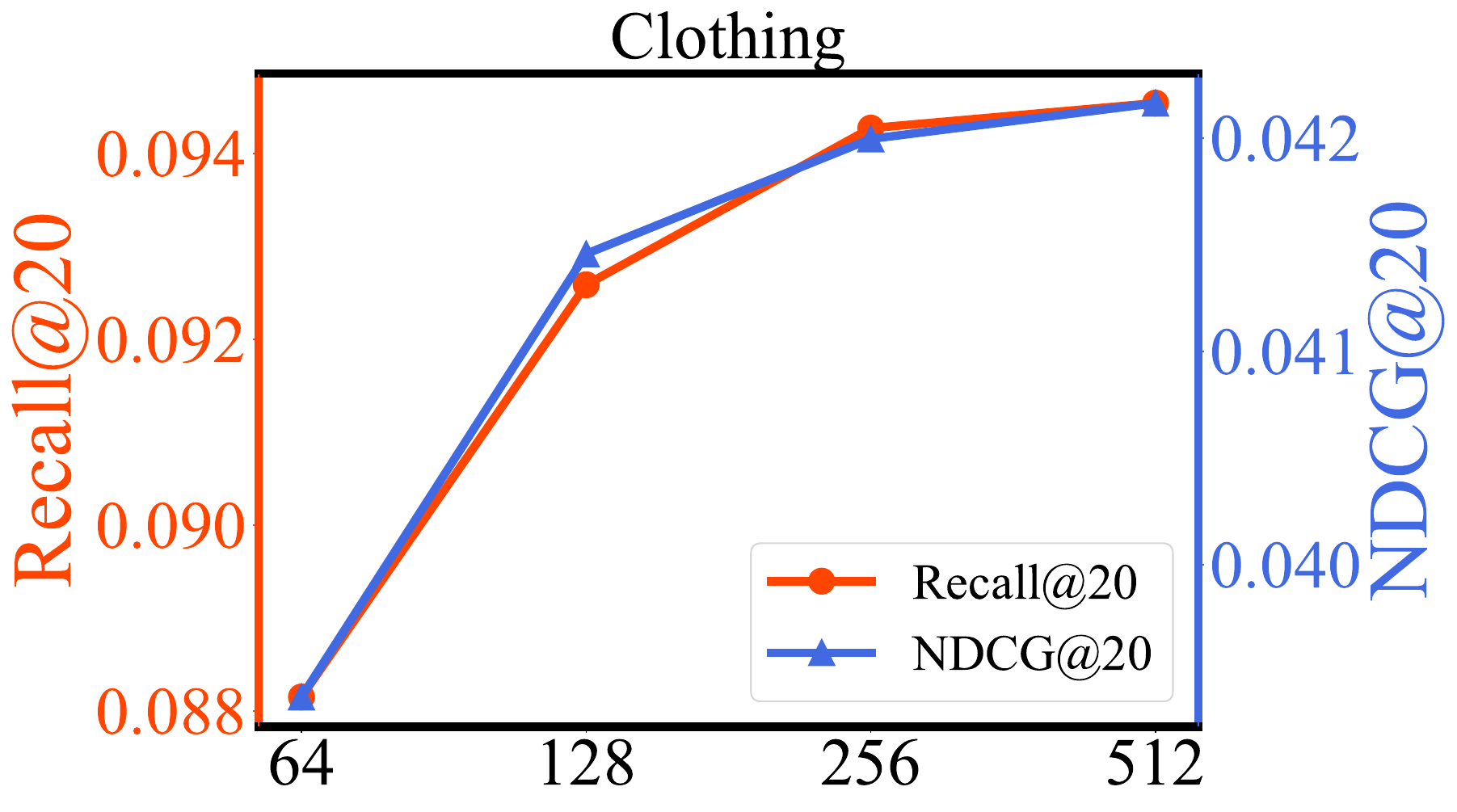}  
}     
\setcounter{subfigure}{2} 
\subfigure[$L_{2}$ Regularization Coefficient $\lambda$] {
\includegraphics[width=0.43\textwidth]{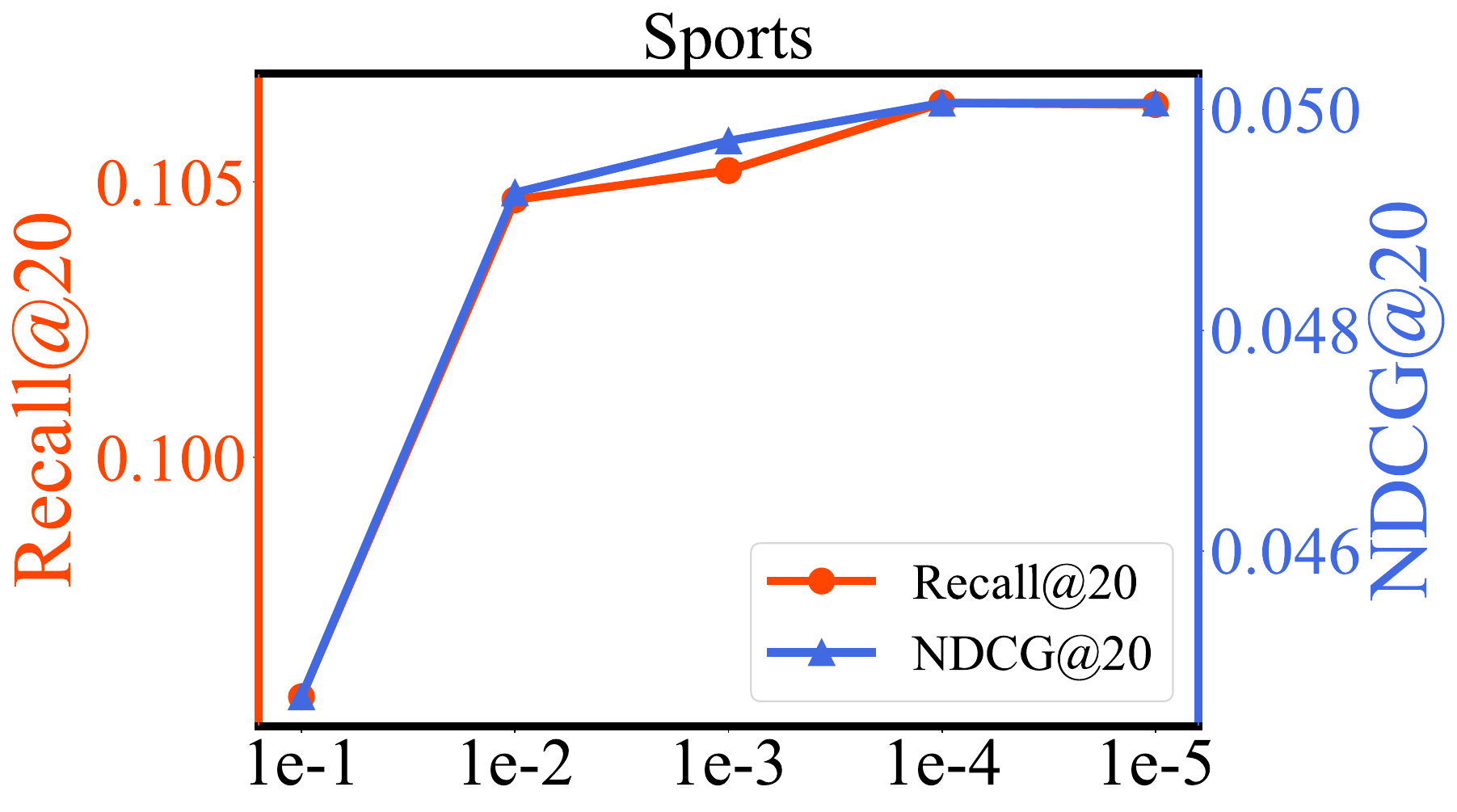}  
}     
\setcounter{subfigure}{5} 
\subfigure[dimension $d$] {
\includegraphics[width=0.43\textwidth]{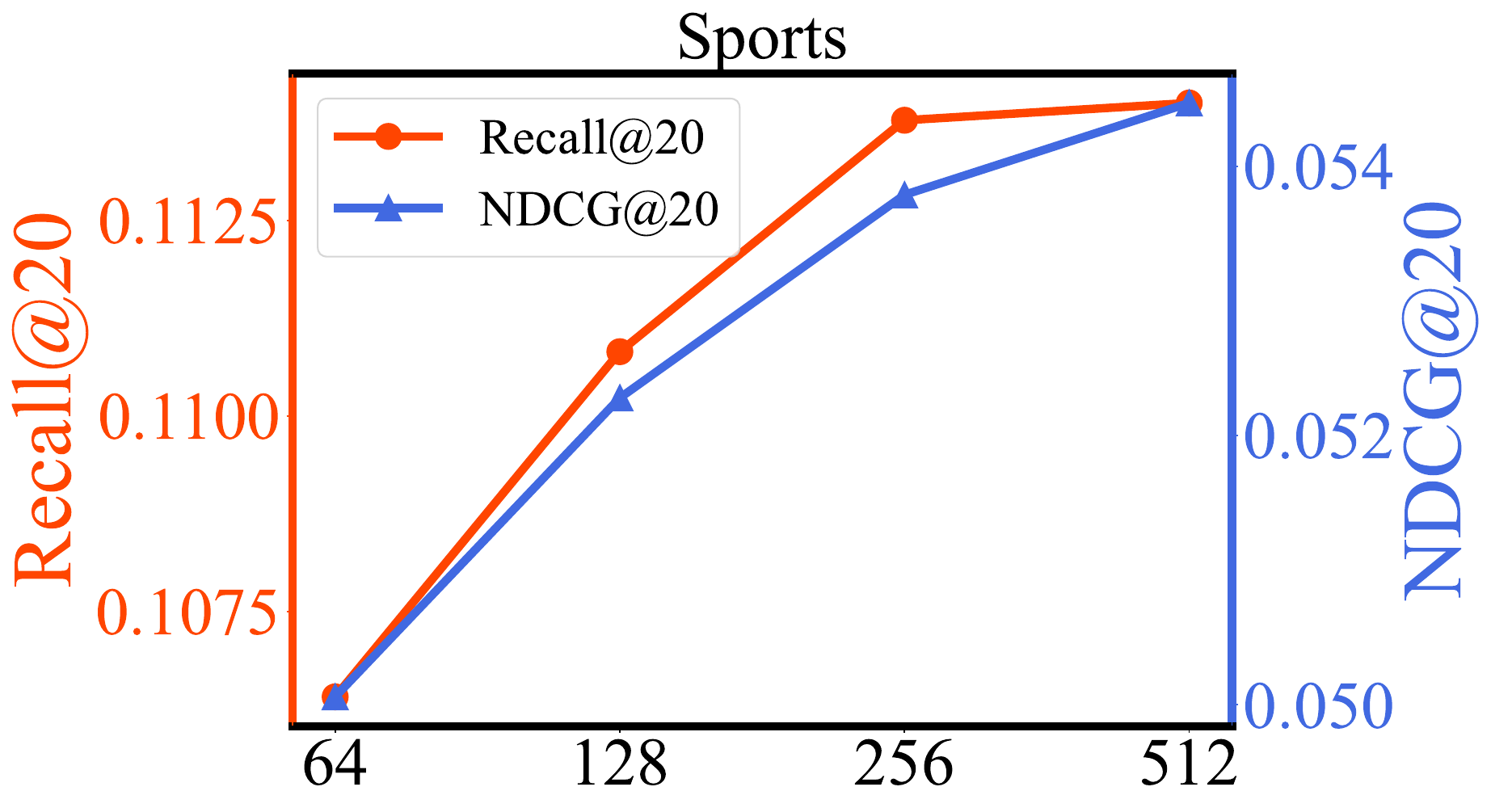}  
}     
\caption{Performance comparison w.r.t various hyperparameters on the three datasets.}
\label{hyperparameters}
\vspace{-5pt}
\end{figure*}

\subsubsection{Impact of $L_{2}$ Regularization Coefficient $\lambda$}
\
    

    The choice of $L_{2}$ regularization coefficient $\lambda$ has a significant influence on the performance of the MMHCL model, particularly in its ability to prevent overfitting. As shown in Fig. \ref{hyperparameters}(a)-(c), excessively large values of $\lambda$ result in over-penalization, leading to accelerated weight decay and a subsequent degradation in performance. This is because larger regularization coefficients impose a heavy constraint on the model parameters, which may hinder the model's capacity to learn complex relationships in the data.

    Conversely, using a properly tuned $L_{2}$ coefficient effectively mitigates overfitting by applying a balanced degree of regularization. Across the TikTok, Clothing, and Sports datasets, we observe that moderate values of $\lambda$ (e.g.,$1 e^{-2}$ or $1 e^{-3}$) strike the right balance between under- and over-regularization, leading to superior performance in terms of both Recall@20 and NDCG@20 metrics. This suggests that regularization is crucial for controlling model complexity and ensuring generalization, particularly in recommendation tasks with complex multimodal data.

\subsubsection{Impact of the different latent dimension $d$}
\

    The latent dimension $d$ of the embeddings plays a critical role in the MMHCL model’s ability to capture and represent user-item relationships effectively. As illustrated in Fig. \ref{hyperparameters}(d)-(f), increasing the embedding dimension results in a significant boost in model performance across all three datasets. This is because larger embedding dimensions enable the model to encapsulate more comprehensive information about users and items, allowing for richer, more nuanced representation learning.

    While a dimension of 64 was used for consistency across comparisons in earlier sections, our findings suggest that higher dimensionality (e.g., 256 or 512) is beneficial for improving recommendation accuracy. Specifically, as the embedding dimension increases, the model gains greater capacity to represent the complex semantic relationships present in multimodal data, resulting in enhanced performance for both Recall@20 and NDCG@20 metrics. However, the potential benefits of further increasing $d$ should be balanced against the increased computational cost and risk of overfitting, especially in scenarios where training data may be limited.
    These results emphasize the importance of carefully tuning the latent dimension to fully leverage the model’s potential in capturing rich semantic structures while maintaining computational efficiency.

\subsubsection{Impact of HGNN Layers}
\

\begin{figure}[t]
	\centering
	\includegraphics[width=0.6\textwidth ]{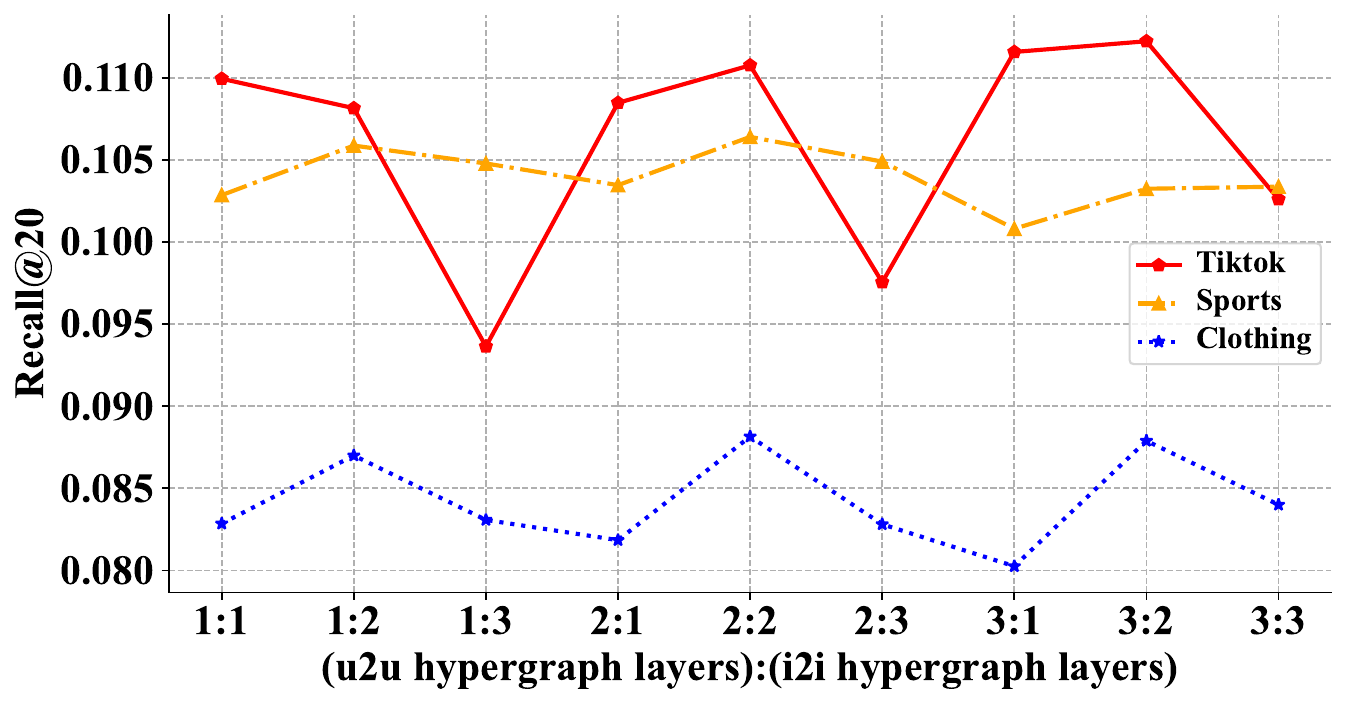}
	\caption{Performance w.r.t the depth of u2u hypergraph layers and i2i hypergraph layers on the three datasets (Recall@20).}
	\label{2.gnn_layers}
\end{figure}

To investigate the impact of hypergraph depth on the MMHCL model’s ability to capture higher-order semantic information, we explored various combinations of user-to-user (u2u) and item-to-item (i2i) hypergraph layers. As illustrated in Fig. \ref{2.gnn_layers}, we evaluated different layer depths, ranging from 1 to 3, for both the u2u and i2i hypergraphs across the three datasets. 

The results demonstrate a general trend where increasing the depth from 1 to 2 layers enhances performance across all datasets. This improvement indicates that a deeper hypergraph structure allows the model to capture richer, higher-order semantic relationships among users and items, which significantly improves the quality of recommendations. In particular, the model benefits from more complex, multi-step interactions as deeper layers enable the network to aggregate information over larger neighborhoods, effectively capturing shared preferences and item similarities.

However, as the depth of the hypergraph continues to increase to 3 layers, performance begins to decline. This reduction can be attributed to over-smoothing, where excessive message passing dilutes the distinctiveness of node embeddings, leading to a loss of critical information. Moreover, deeper networks may introduce irrelevant relations and noise, particularly in dense hypergraph structures, which negatively impacts the model’s ability to learn meaningful interactions. 

These findings suggest that while deeper hypergraphs are beneficial for capturing more comprehensive semantic information, there is a limit to the depth beyond which performance declines due to over-smoothing and noise. Therefore, careful tuning of hypergraph depth is essential for maximizing the model’s performance, particularly in complex multimodal recommendation tasks.


\subsection{Exploration of Modalities}
\

\begin{figure}[tbp]
    \centering
    \includegraphics[width=0.45\textwidth]{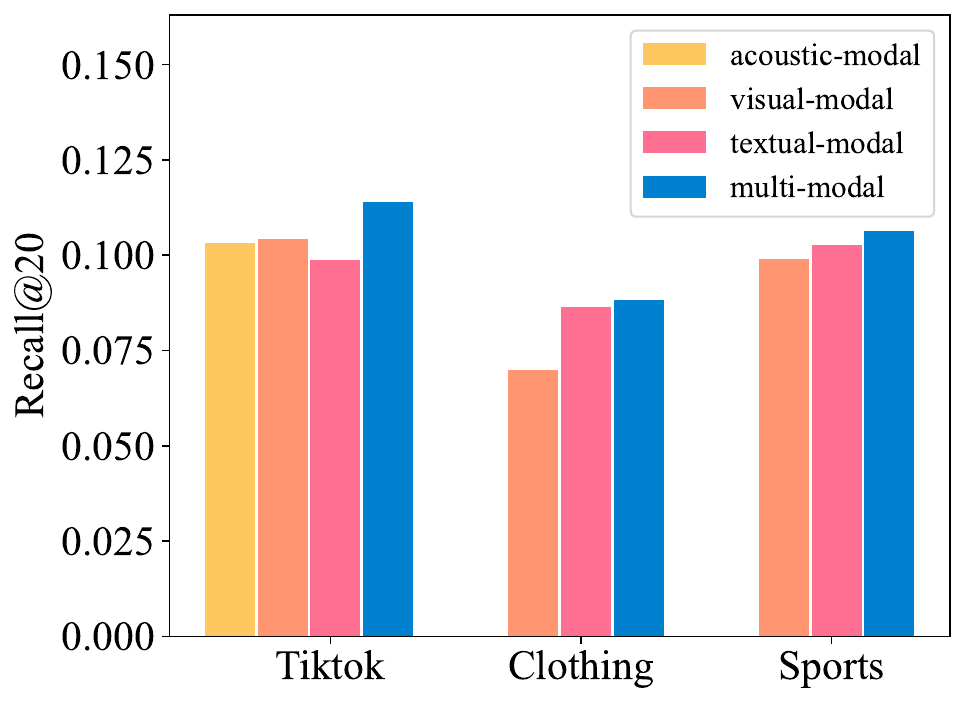}
    \caption{Performance w.r.t the effect of different modalities on the three datasets (Recall@20). }
    \label{modal}
    
\end{figure}
    As shown in Fig. \ref{modal}, we evaluate the effects of different modalities on the TikTok, Clothing, and Sports datasets. The results consistently demonstrate that multimodal approaches outperform unimodal ones, highlighting the importance of integrating various data sources to enhance recommendation accuracy. 

In the TikTok dataset, where the content is predominantly short-form video, visual and acoustic modalities are key drivers of performance. TikTok videos inherently emphasize visual appeal and sound quality, which provide rich and immediate cues about user preferences. For instance, visual features like color schemes, motion patterns, and facial expressions, along with audio elements such as background music or voiceovers, play an essential role in engaging users and guiding their interactions. The results highlight that, in this context, models incorporating both visual and acoustic information yield more accurate recommendations, as they capture the core elements driving user engagement on the platform. 

Conversely, in the Amazon Clothing and Sports datasets, textual data proves to be more influential than visual information. This finding is intuitive, as users shopping for clothing or sports equipment often rely heavily on product descriptions, reviews, and specifications, which provide critical information about the items’ features, such as size, material, or functionality. While images offer a visual reference, they cannot fully convey the detailed attributes users need to make informed purchasing decisions. As a result, models that emphasize textual information—such as descriptions, brand, and title—outperform those that primarily focus on visual features in these domains. 

    Our model demonstrates strong performance in fusing multiple modalities, effectively leveraging complementary information from visual, textual, and acoustic sources. This multimodal fusion allows the model to capture a more holistic view of both users and items, leading to more accurate recommendations. Furthermore, the model’s flexibility across domains, from video-driven engagement in TikTok to text-driven decisions in Clothing and Sports, showcases its adaptability and robustness in diverse recommendation scenarios. 
    
\section{Conclusion and Future Work}

    In this work, we propose a novel framework, Multimodal Hypergraph Contrastive Learning (MMHCL), designed to enhance recommendation systems by leveraging hypergraphs to capture complex second-order semantic correlations among users and items. By modeling shared item preferences among users and multimodal similarities between items, MMHCL enables a deeper understanding of both user behavior and item relationships. The introduction of a contrastive feature enhancement paradigm, which applies synergistic contrastive learning between first-order and second-order embeddings, further improves feature distinguishability and robustness.
Our extensive experimental evaluation demonstrates that MMHCL not only mitigates common issues such as data sparsity and cold-start problems but also excels in multimodal fusion, effectively utilizing complementary information across different modalities. These results underscore the strength of our approach in addressing the unique challenges of recommendation systems and provide a promising direction for future research in multimodal learning. 

    Despite its strengths, our method has some limitations similar to prior approaches, suggesting avenues for future research. 
 Specifically, MMHCL relies on the availability and quality of multimodal content features. If certain items have missing or poor-quality modality data (e.g., low-quality images or very sparse text), the benefits of the i2i hypergraph may diminish. In such cases, integrating advanced pre-trained models to generate better item representations (for example, using large vision-language models or diffusion models to enrich item content) could enhance performance. 
The impact of product descriptions across different countries and languages, as well as the effective utilization of multimodal features, remains an open research problem. 
    
\bibliographystyle{ACM-Reference-Format}
\bibliography{sample-base}


\end{document}